\newif\ifcameraready
 \camerareadytrue

\newif\ifonlineversion
\onlineversionfalse

\newcommand{\versionnum}[0]{3.0}

\documentclass[10pt, conference, compsoc]{IEEEtran}
\usepackage{mathptmx} 

\usepackage{authblk}
\usepackage{fancyhdr}
\usepackage[normalem]{ulem}
\usepackage[hyphens]{url}
\usepackage[sort,nocompress]{cite}
\usepackage[final]{microtype}
\usepackage[keeplastbox]{flushend}
\usepackage[dvipsnames]{xcolor}
\usepackage{footmisc}

\usepackage{booktabs} 
\usepackage{xspace}
\usepackage{enumitem}
\usepackage[binary-units=true]{siunitx}
\usepackage{amsmath}
\usepackage{ifthen}
\usepackage{graphicx}
\usepackage{multirow}
\usepackage{listings}
\usepackage{setspace}
\usepackage{float}
\usepackage[us,12hr]{datetime}
\usepackage[hyphens]{url}
\usepackage{tikz}
\usepackage{dblfloatfix}
\usepackage{nowidow}

\usepackage[bookmarks=true,breaklinks=true,letterpaper=true,colorlinks=false,hidelinks]{hyperref}

\ifcameraready
\ifonlineversion
\else
\fi
\else
\fancypagestyle{firstpage}
{
  \fancyhead{}
  \fancyhead[C]{\textcolor{MidnightBlue}{\emph{Version \versionnum~---~\today, \ampmtime}}}
}

\fi

\newcommand{\paperlinespacing}[0]{0.81}

\DeclareSIUnit{\flop}{FLOP}

\newcommand{\titleShort}{Mensa\xspace}
\newcommand{\exampleDesign}{\titleShort-G\xspace}
\newcommand{\accelA}{Pascal\xspace}
\newcommand{\accelB}{Pavlov\xspace}
\newcommand{\accelC}{Jacquard\xspace}

\newcommand{\paratitle}[1]{\vspace{4pt}\noindent\textbf{#1.}}
\newcommand{\paratitleattop}[0]{\vspace{-4pt}}
\newcommand{\paratitlespacing}[0]{\vspace{4pt}\noindent}

\newcommand{\todob}[1]{}
\newcommand{\am}[1]{#1}
\newcommand{\sg}[1]{{#1}}
\newcommand{\sgii}[1]{{#1}}

\newcommand{\shep}[1]{#1}
\newcommand{\shepii}[1]{#1}

\ifcameraready
  \newcommand{\todo}[1]{}
  \newcommand{\revi}[1]{#1}
  
  \newcommand{\gfrevii}[1]{#1}
  \newcommand{\revii}[1]{#1}

  \newcommand{\om}[1]{}
  \newcommand{\gf}[1]{}
\else
  \newcommand{\todo}[1]{{\bf\color{Red} TODO: #1}}
  \newcommand{\revi}[1]{{\color{MidnightBlue} #1}}
  
  \newcommand{\gfrevii}[1]{{\color{Orange} #1}}

  \newcommand{\revii}[1]{{\color{Green} #1}}
  \newcommand{\om}[1]{{\color{BrickRed}\bf (OM: #1)}}
  \newcommand{\gf}[1]{{\color{orange}\bf (GF: #1)}}

\fi

\makeatletter
\g@addto@macro{\normalsize}{%
  \setlength{\abovedisplayskip}{2pt plus 0.5pt minus 1pt}
  \setlength{\belowdisplayskip}{2pt plus 0.5pt minus 1pt}
  \setlength{\abovedisplayshortskip}{0pt}
  \setlength{\belowdisplayshortskip}{0pt}
  \setlength{\intextsep}{2pt plus 1pt minus 1pt}
  \setlength{\textfloatsep}{2pt plus 1pt minus 1pt}
  \setlength{\skip\footins}{2pt plus 1pt minus 1pt}}
\makeatother

\title{\vspace{-16pt}Google Neural Network Models for Edge Devices:\\ Analyzing and Mitigating Machine Learning Inference Bottlenecks}

\newcommand{\affilCMU}[0]{$^\dagger$}
\newcommand{\affilETH}[0]{$^\star$}
\newcommand{\affilUIUC}[0]{$^\ddagger$}
\newcommand{\affilStanford}[0]{$^\diamond$}
\newcommand{\affilGoogle}[0]{$^\S$}

\author{\vspace{-30pt}\\%
Amirali Boroumand\affilCMU\affilStanford%
\quad\qquad%
Saugata Ghose\affilUIUC%
\quad\qquad%
Berkin Akin\affilGoogle%
\quad\qquad%
Ravi Narayanaswami\affilGoogle%
\\%
Geraldo F. Oliveira\affilETH%
\quad\qquad%
Xiaoyu Ma\affilGoogle%
\quad\qquad%
Eric Shiu\affilGoogle%
\quad\qquad%
Onur Mutlu\affilETH\affilCMU%
}

\affil{\normalsize%
\vspace{-15pt}\\%
{\affilCMU}\emph{Carnegie Mellon Univ.}%
\qquad%
{\affilStanford}\emph{Stanford Univ.}%
\qquad%
{\affilUIUC}\emph{Univ.\ of Illinois Urbana-Champaign}%
\qquad
{\affilGoogle}\emph{Google}%
\qquad%
{\affilETH}\emph{ETH Z{\"u}rich}%
\vspace{-10pt}}

\begin{document}
\sloppy
\maketitle
\ifcameraready
\else
  \thispagestyle{firstpage}
\fi
\pagestyle{plain}


\setstretch{\paperlinespacing}
\renewcommand{\footnotelayout}{\setstretch{\paperlinespacing}}

\begin{abstract}

\revi{Emerging edge computing platforms often contain machine learning (ML) accelerators
that can accelerate inference for a
wide range of neural network (NN) models\revii{. These models} are designed to fit
within the limited area and energy constraints of the \revii{edge computing} platforms\revii{, each targeting various applications (e.g., face detection, speech recognition, translation, image captioning, video analytics)}.} \revi{To understand how edge ML accelerators perform, 
we characterize the performance of a commercial Google Edge TPU,
using 24 Google edge NN models (which span a wide range of NN model types)
and analyzing each NN layer within each model.
We find that the Edge TPU suffers} from three \revii{major} shortcomings:
\revi{(1)~it operates significantly below peak computational throughput,
(2)~it operates significantly below its theoretical energy efficiency, and
(3)~\revii{its memory system is a large energy and performance bottleneck}.} \revi{Our characterization reveals that the one-size-fits-all\revii{, monolithic} design of the Edge TPU
ignores the high degree of heterogeneity both across different NN models and across different NN layers within the same NN model,
leading to the shortcomings we observe.}

\sg{We} propose a new \sg{acceleration} framework called Mensa. Mensa incorporates multiple heterogeneous \revi{edge ML} accelerators \am{(including both on-chip and near-data accelerators)}, each of which caters to the characteristics of a particular subset of \revi{NN} models \revii{and layers}. 
\revi{\revii{During NN inference}, for each NN layer, Mensa decides which accelerator to schedule the layer on,
taking into account both the optimality of each accelerator for the layer and layer-to-layer communication costs.} \sg{\revii{Our comprehensive analysis of} the Google edge NN models} \revii{shows} that all of the layers naturally group into a small number of clusters, 
\sg{which allows us to design an efficient implementation of Mensa for these models with only three specialized accelerators.} \am{Averaged across all 24 Google edge \revi{NN} models, Mensa improves energy efficiency and throughput by 3.0x and 3.1x
over the Edge TPU, and by 2.4x and 4.3x over Eyeriss~v2, a state-of-the-art accelerator}.
\end{abstract}


\section{Introduction}  
\label{sec:intro}

Modern consumer devices make widespread use of
machine learning (ML).
The growing complexity of these devices, combined with 
increasing demand for privacy, connectivity, and real-time responses,
has spurred significant interest in pushing ML inference computation to
the edge (i.e., \revi{in or near these consumer} devices, instead of the cloud)~\cite{edge-facebook, edge-nature,eyerissv2}.
Due to \revi{their resource-constrained nature, edge \revii{computing} platforms}
now employ specialized
energy-efficient accelerators for on-device inference
(e.g., Google Edge \revi{Tensor Processing Unit, TPU~\cite{edge-tpu};} NVIDIA Jetson~\cite{jetson};
Intel Movidius~\cite{movidius}).
At the same time, neural network (NN) algorithms are evolving rapidly,
which has led to \sg{many types of} NN models
(e.g., \revi{convolutional neural networks, CNNs~\gfrevii{\cite{fukushima.biologicalcybernetics1980, lecun.cognitiva1985, rumelhart.nature1986, lecun.nature2015, simonyan2015very,gu2018recent,lecun1989handwritten,lecun1998gradient,russakovsky2015imagenet,zeiler2014visualizing,szegedy2015going,he.cvpr2016}};
long short-term memories, LSTMs~\gfrevii{\cite{hochreiter.neco1997, gers.icann1999, greff.tnnls2017,lstm-google,graves2013generating,google-translation,cho2014learning, lrcn, karpathy2015deep,ranzato2014video,srivastava2015unsupervised,sutskever2014sequence,xu2015show,xingjian2015convolutional}};
gated recurrent units, GRUs~\gfrevii{\cite{gru,kanai2017preventing,cho2014properties}}; Transducers~\cite{graves.icmlworkshop2012, he.icassp2019};
hybrid models such as recurrent CNNs, RCNNs~\cite{liang.cvpr2015, pinheiro.icml2014, lrcn,rcnn-google}}), 
each targeting various applications 
(e.g., face detection~\gfrevii{\cite{simonyan2015very,jiang2017face,qin2016joint,li2015convolutional}},
speech recognition~\gfrevii{\cite{he.icassp2019,transducer3, lstm-google,graves2013hybrid,han2017ese,soltau2016neural}}, 
translation~\cite{google-translation}, 
image captioning~\cite{rcnn-google,lrcn}\revii{, video analytics~\cite{hsieh2018focus}}).

\revi{At a high level, all of these different types of
NN models consist of multiple \emph{layers},
where each layer takes in a series of
\emph{parameters} (i.e., weights) and
\emph{input activations} (i.e., input data),
and produces a series of
\emph{output activations} (i.e., output data).
As a result, despite the wide variety of NN model types,}
\sg{Google's state-of-the-art Edge TPU~\cite{edge-tpu} provides an optimized 
one-size-fits-all \revi{design (i.e., a monolithic accelerator with a fixed\revii{, large} 
number of processing elements and a fixed \emph{dataflow},
which determines how data moves between
the accelerator's components)}
\revi{that} \revii{caters to} edge device area
and energy constraints.}
Unfortunately, \sg{we find that it is very challenging to simultaneously achieve}
high energy efficiency (\si{\tera\flop\per\joule}),
computational throughput (\si{\tera\flop\per\second}), and
area efficiency (\si{\tera\flop\per\milli\meter\squared})
for each \revi{NN model} \sg{with this one-size-fits-all \revi{design}}. We conduct an in-depth analysis of \revi{ML} inference execution on \am{a commercial Edge TPU}, across 24 state-of-the-art Google edge \revi{NN} models spanning four popular NN model types:
(1)~CNNs, 
(2)~LSTMs~\cite{lstm-google}, 
(3)~Transducers~\cite{he.icassp2019, transducer3, transducer4}, and 
(4)~RCNNs~\cite{rcnn-google,lrcn}.
\sg{These models are used in several Google mobile applications,
such as image classification, object detection, semantic segmentation,
automatic speech recognition, and image captioning.}
Based on our analysis (Section~3), we find that the \revi{Edge TPU} suffers from
 three major shortcomings. First, 
\sg{the \revi{Edge TPU} utilizes only \revii{24\%} of its peak throughput,
averaged across all models (less than 1\%
for LSTMs and Transducers \revii{in} the worst case)}. Second, despite using specialized logic, the \revi{Edge TPU} \sg{provides only 37\% of} its theoretical peak energy efficiency (\si{\tera\flop\per\joule}) on average \revii{(34\% in the worst case, 51\% in the best case)}. Third, the \revi{Edge TPU}'s memory system is often \sg{a large bottleneck}.
As an example, while large on-chip storage buffers
(e.g., several megabytes) account for a significant portion of \revi{overall} energy consumption
(e.g., 48.1\% static and 36.5\% dynamic energy during CNN inference),
they are often ineffective \sg{at} reducing off-chip accesses, and cannot accommodate
the parameters of larger NN models, \revi{leading to significant energy waste}.

To identify the root cause of these shortcomings, we perform \sg{the first} comprehensive 
per-layer analysis of \sg{the Google edge NN models} \revii{revealing} two key observations.
First, there is significant variation in terms of
 layer type, shape, and characteristics (e.g., \revii{\si{\flop\per\byte}} ratio, \revii{memory} footprint, intra- and inter-layer
 dependencies) \emph{across} \revi{different types of} models. 
For example, Transducer layers differ drastically (by as much as two orders 
of magnitude) from CNN layers in terms of parameter footprint and \si{\flop\per\byte}.
Second, even \emph{within each model}, there is high variation in terms of 
layer types and shapes 
(e.g., pointwise, depthwise, fully-connected, standard convolution, recurrent).
This leads to up to two orders of magnitude of variation for
layer characteristics within a single model.
We \sgii{quantify for the first time how} intra-model \sg{variation} is dramatically higher in
edge models \sg{compared to previously-studied}
traditional models (e.g., \cite{alex-net,simonyan2015very}), as edge models employ
several techniques (e.g., \revii{depthwise separable convolutions~\cite{squeezenet,mobilenet}, pointwise group convolution~\cite{gibson2020optimizing}, channel shuffle~\cite{zhang2018shufflenet,shufflenetv2}})
to reduce computational complexity and layer footprint,
in order \sg{to optimize the models for} resource-constrained edge devices.

\sg{Despite this large variation \revi{across and within NN models}, many} state-of-the-art edge ML
 accelerators \revi{(e.g., \cite{edge-tpu, movidius, movidius-arch, jetson, tangram,brainwave})} take a monolithic\revii{, one-size-fits-all} design approach, where they
 equip the accelerator with a large \revii{fixed-size} \revi{processing element (PE)} array, large on-chip buffers, and a fixed
 dataflow (e.g., output stationary). While this approach might \revi{lead to efficient execution} for
 a specific \revii{family} of layers (e.g., traditional convolutional layers with high \revi{computational} intensity
 and high data reuse), we find that it 
\sg{leads to \revii{large} throughput and energy efficiency shortcomings} across the \sg{significantly more} diverse edge NN models,
\sg{as illustrated by two examples}. \revi{First, despite the existence of large on-chip buffers
in state-of-the-art accelerators, many edge NN models
still \revi{cannot fit all of their parameters in the buffers and} generate a large amount of off-chip memory traffic,
and the resulting \revii{memory} bandwidth bottlenecks lead to 
PE underutilization.} \sg{Second, state-of-the-art accelerators use a fixed dataflow across all layers.} Due to the drastic variation \revii{in layer characteristics} across different layers,
the fixed dataflow often misses spatial/temporal reuse opportunities across layers.

\sg{A number of recent works\revi{~\cite{eyerissv2,maeri,scaledeep,kwon.hpca2021}} cater to NN variation by enabling reconfigurability for
\revi{a subset of the accelerator components}.  For example, Eyeriss~v2~\cite{eyerissv2} provides 
the ability to reconfigure the on-chip interconnect and make use of a smaller
PE array.  Unfortunately, as models become more diverse and go beyond the
structure of more traditional CNNs, \revi{existing} reconfigurable accelerators face \revi{two key} issues:
(1)~they do not provide the ability to reconfigure \revii{or customize} a number of essential
design parameters (e.g., on-chip buffers, memory bandwidth),
\revi{which makes it very difficult to co-optimize the dataflow with the memory system; and}
(2)~they can require frequent \sg{online} reconfiguration to cater to increasing intra-model
heterogeneity, \sg{with associated overheads}.} \am{The \emph{key takeaway} from our \sg{extensive analysis of Google edge NN models on the Edge TPU}
is that \emph{all key components} 
of an edge accelerator (i.e., PE array, dataflow, memory system)
must be co-designed \revii{and co-customized} based on specific layer characteristics to achieve high utilization and energy efficiency.}
Our goal is to revisit the design of edge ML accelerators such that they are aware of and can \sg{fully} exploit
the \sg{growing} variation within and across edge NN models.

\am{To this end, 
we} propose \titleShort, the first general HW/SW composable
framework for ML acceleration in edge \revii{computation} devices.
\sg{The key idea of \titleShort is to \revii{perform NN} layer
execution across \revii{\emph{multiple}} on-chip and near-data accelerators,
each of which is small and tailored to the characteristics of a particular
subset \revii{(i.e., family)} of layers.
\revii{Our rigorous experimental study of} the characteristics of different layers in \revi{the Google edge NN} models \revii{reveals} that the layers naturally group into a small number of
clusters that are based on a subset of these characteristics.
This \revii{new insight} allows us to \revi{significantly} limit the number of different accelerators 
\revi{required} in a \titleShort design. We design a runtime \revi{scheduler} for \titleShort to determine which of these
accelerators should execute \revi{which \revii{NN} layer},
using information about
(1)~which accelerator is
best suited to the layer's characteristics, and 
(2)~inter-layer dependencies.}

\sg{Using our \revii{new} insight about layer clustering, we develop 
\revi{\emph{\exampleDesign}, an example} design for \titleShort that is
optimized for \revi{Google edge NN models}.
We find that \revi{the design of \exampleDesign's underlying accelerators} should center around two
key layer characteristics (memory boundedness, and 
activation/parameter reuse opportunities).
\revi{This allows us to provide efficient inference execution
for \revii{\emph{all}} of the Google edge NN models using \revii{\emph{only
three}} accelerators in \exampleDesign
(we call the individual accelerators \accelA, \accelB, and \accelC).}} \revi{\accelA, for compute-centric layers,
maintains the high PE utilization that these layers
achieve in the Edge TPU, but does so using an
optimized dataflow that both reduces the size of
the on-chip buffer (16x smaller than in the Edge TPU)
and the amount of on-chip network traffic.
\accelB, for LSTM-like data-centric layers,
employs a dataflow that enables the temporal reduction
of output activations, and
enables the parallel execution of layer operations in
a way that increases parameter reuse,
\revii{greatly} reducing off-chip memory traffic.
\accelC, for other data-centric layers,
significantly reduces the size of the on-chip parameter buffer
\revii{(by 32x)}
\revii{using} a dataflow that exposes reuse opportunities
for parameters.
As both \accelB and \accelC are optimized for data-centric layers,
which require significant \revii{memory} bandwidth and are unable to
utilize a significant fraction of PEs in the Edge TPU,
we place the accelerators in the logic layer of 3D-stacked
memories and use significantly smaller PE arrays
compared to the PE array in \accelA,
\revii{unleashing} significant performance \emph{and} energy benefits.}

Our evaluation shows that compared to \am{\revii{the} baseline Edge TPU},
\revi{\exampleDesign} reduces total inference energy by 66.0\%,
 improves energy efficiency (\si{\tera\flop\per\joule}) by 3.0x, and increases computational throughput (\si{\tera\flop\per\second}) by 3.1x,
\revii{averaged across all 24 Google edge NN models}. \am{\revi{\exampleDesign} improves inference energy 
efficiency and throughput 
by 2.4x and 4.3x over Eyeriss~v2, a state-of-the-art accelerator}.

\paratitlespacing{}
We make the following \textbf{contributions} in this work:

\begin{itemize}[leftmargin=1em,nosep]

\item We conduct the first in-depth analysis of how \sg{the Google Edge TPU operates}
across a wide range of state-of-the-art \sg{Google} edge NN models.
Our analysis reveals three key shortcomings of \am{the Edge TPU}
\sg{for these models}:
(1)~\sg{poor throughput due to} significant PE underutilization,
\sg{(2)~\am{low energy efficiency}, and
(3)~a large memory bottleneck.}

\item We comprehensively analyze the key characteristics of each layer in 
\sg{24 Google} edge NN models. We make \sgii{three} observations from our analysis:
(1)~layer characteristics vary significantly both \revi{\emph{across models
and across layers within a single model}},
(2)~the monolithic, \revii{one-size-fits-all} design of \am{state-of-the-art accelerators (e.g., the Edge TPU)} is the
root cause of \revi{\revii{shortcomings} for edge ML inference}, and
\sgii{(3)~layers naturally group into a small number of clusters based on their
characteristics}.

\item We propose \titleShort, a new framework for efficient edge ML acceleration.
\sg{\titleShort is the first \revii{ML} accelerator to exploit the significant compute and \revi{memory}
heterogeneity that we observe in state-of-the-art edge NN models,
through the use of a few small, \revi{carefully-specialized}
accelerators, \revii{a runtime scheduler to orchestrate layer execution on the heterogeneous accelerators}.}

\item
We \sg{create \revi{\exampleDesign, an example} \titleShort design for our Google edge models. \revii{We} find that, \revii{with its three specialized accelerators, \exampleDesign}} is significantly more energy efficient and \revi{higher performance}
than \am{a commerical Edge TPU} \sg{and Eyeriss v2, a state-of-the-art ML accelerator.} 
\end{itemize}


\section{Background}
\label{sec:bkgd}

\revi{We provide a brief background on the four major types of
neural network (NN) models that we evaluate in this work.
Detailed descriptions can be found in other works~\cite{lecun.nature2015, greff.tnnls2017, graves.icmlworkshop2012, lrcn}.}

\paratitle{CNNs}
\revi{Convolutional neural networks (CNNs)~\cite{fukushima.biologicalcybernetics1980, lecun.cognitiva1985, rumelhart.nature1986, lecun.nature2015, simonyan2015very}
are feed-forward multi-layer models
that are designed to capture spatial features.
CNNs are typically used for applications such as
image classification and object detection,
where the identification of a visual feature is required~\cite{lecun.nature2015, simonyan2015very}.
A CNN is composed mainly of convolutional layers,
which are used to downsample the input and detect different features.
Each convolutional layer
(1)~performs a 2D convolution operation between 
the \emph{input activations} (e.g., a slice of an image, downsampled features)
and \emph{parameters} (i.e., the weights for that layer), 
where the parameters consist of one or more \emph{kernels}
(small matrices that apply an operation to a small portion of the input; 
e.g., sharpening a slice of an image);
and (2)~passes the result
through a non-linear activation function (e.g., ReLU, sigmoid, tanh)
to produce the \emph{output activations} (the detected features). 
At the end of the model, fully-connected layers
combine the features generated from different
convolutional layers to perform the final classification.
but can also include fully-connected, depthwise, and pointwise layers.
A CNN typically takes in some spatially-oriented input
(e.g., image, video) and returns a classification.}

\paratitle{LSTMs} 
\revi{Long short-term memory (LSTM) networks~\cite{hochreiter.neco1997, gers.icann1999, greff.tnnls2017}
are multi-layer models
with recurrent connections (i.e., data from one iteration of
a layer is reused in a subsequent iteration of the same layer)
that are effective at classifying sequences (i.e., ordering) of data
and predicting future sequences.
LSTMs are used for applications such as traffic forecasting~\cite{zhao.ietits2017}, text reply prediction~\cite{kannan.kdd2016}, and handwriting recognition~\cite{greff.tnnls2017}.
An LSTM network consists of multiple LSTM layers,
each of which includes several LSTM cells.
Within each LSTM cell, there are four \emph{gates}
(input, input modulation, forget, and output)\footnote{\revi{The input gate and input modulation gate are sometimes collectively referred to as the input gate.}} that
allow the cell to regulate information flow and update
the state of the cell accordingly.
At each iteration, a cell makes a prediction based on 
the current input ($x_{t}$) and the \emph{hidden vector} ($h_{t-1}$), 
which consists of the activations from the previous time step and
serves as the recurrent connection.
Each gate performs two matrix-vector \revi{multiplications} (MVM):
(1)~an input MVM \revi{of} the input parameter matrix ($W_{x}$)
and the input vector ($x_{t}$), and 
(2)~a hidden MVM \revi{of} the hidden parameter matrix ($W_{h}$)
and the input hidden vector ($h_{t-1}$). An LSTM network takes in a sequence of inputs, and returns
a prediction for the entire sequence.}

\paratitle{Transducers}
\revi{Transducers~\cite{graves.icmlworkshop2012, he.icassp2019} are multi-layer recurrent NNs
that are effective at classifying sequences of data while
being invariant to distortions or variations in the input data.
Transducers are used for applications such as 
automatic speech recognition~\cite{graves.icmlworkshop2012, he.icassp2019, transducer3,transducer4}.
A transducer has three major \revi{components}:
(1)~an encoder, which receives acoustic features and converts
them into a high-level representation;
(2)~a prediction network, which generates linguistic outputs (i.e., high-level representation) 
that depend on the entire sequence of labels; and
(3)~a joint, which is a feed-forward joint that receives inputs from
both the encoder and a prediction network that depends only on
label histories.  
Each of these components is typically implemented by stacking several LSTM layers.
Like an LSTM network, a Transducer takes in a sequence of inputs,
and returns a prediction for the entire sequence.}

\paratitle{RCNNs}
\revi{Recurrent convolutional neural networks (RCNNs)~\cite{liang.cvpr2015, pinheiro.icml2014, lrcn}
are hybrid multi-layer recurrent NNs that are designed to
capture spatio-temporal information~\cite{rcnn-google, rcnn-caption, dnpu,lrcn-fpga,crnn}.
RCNNs are used for applications such as image captioning~\cite{rcnn-google, rcnn-caption}, 
activity/gesture recognition~\cite{lrcn, dnpu},
video scene labeling~\cite{lrcn, lrcn-fpga},
weather forecasting~\cite{rcnn-weather}, and sound classification~\cite{crnn, crnn2}.
In this paper, we focus on long-term recurrent convolutional networks (LRCNs)~\cite{lrcn}, a popular type of RCNN that typically employs multiple convolutional layers in the front end of the network to perform
spatial feature extraction on input data, 
and then passes the spatial features to an LSTM-based model that
predicts a temporal sequence.
RCNNs take in a sequence of spatially-oriented inputs,
and returns a sequence prediction.}


\section{TPU \& Model Characterization}
\label{sec:motiv}
\label{sec:motiv:methodology}

\shepii{We analyze the performance and energy of executing edge NN models}
using \am{a commercial Edge TPU\revi{~\cite{edge-tpu}} as our baseline accelerator}. \am{The Edge TPU has a generic tiled architecture, similar to other state-of-the-art accelerators~\cite{eyeriss, tetris, tangram, tpu}.
It includes a 2D array of PEs (64x64), where each PE has a small register file to hold intermediate results. 
 The accelerator has two large SRAM-based on-chip buffers to
 hold model parameters and activations~\cite{edge-tpu-compiler}.} \shepii{In our study, we analyze 24 Google edge models (including CNNs, LSTMs, Transducers, and RCNNs).
While we are unable to disclose model specifics, we expect to see similar
performance and energy characteristics for popular publicly-available models such as 
MobileNet~\cite{mobilenet} and ResNet~\cite{he.cvpr2016},
as these public models have similarities to some of the Google models.}

\revi{Our analysis consists of two parts.
First, we the current shortcomings of the Edge TPU (Section~\ref{sec:motiv:accelerator}).
Second, we analyze each edge NN model at the granularity
of layers, to better understand the sources of the Edge TPU shortcomings
(Section~\ref{sec:motiv:model}).
We summarize key takeaways in Section~\ref{sec:motiv:takeaways}.}

\subsection{Google Edge TPU Shortcomings}
\label{sec:motiv:accelerator}

\revi{Based on our analysis, we 
find that the accelerator suffers from three major shortcomings
(we discuss the causes of each shortcoming in Section~\ref{sec:motiv:discussion}):}

\begin{figure}[h]
    \centering
    \centering
    \includegraphics[width=\linewidth]{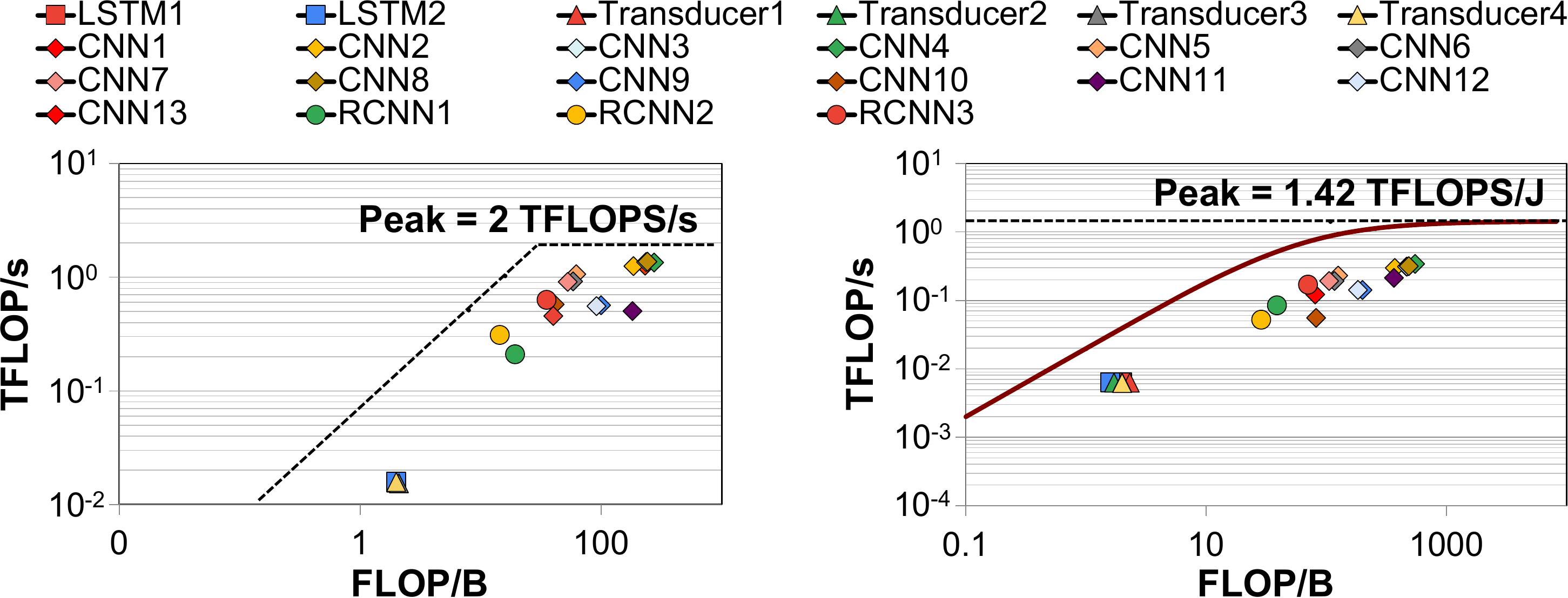}%
    \caption{\revi{Throughput roofline (left) and energy roofline (right) for \am{the Edge TPU} across all \revi{Google edge neural network} models.}}
    \label{fig:roofline-throughput}
    \label{fig:roofline-energy}
\end{figure}

\textbf{1. The accelerator often suffers from extreme underutilization of the PEs.} The \am{Edge TPU} has a theoretical peak throughput of \revi{\SI{2}{\tera\flop\per\second}.} However, 
the accelerator operates \emph{much} lower than
 peak throughput during inference execution (75.6\% lower on average). Figure~\ref{fig:roofline-throughput} (left) shows the
 roofline model of throughput for \am{the Edge TPU},
along with the \sg{measured} throughput of all of our edge models. 
The PE utilization is consistently low across all models.
Transducer and LSTM models have the most underutilization,
with both achieving less than 1\% of peak throughput.
While CNN and RCNN models do somewhat better,
they achieve only 40.7\% of peak utilization on average
(\revi{with a minimum of only 10.2\% of peak utilization}).

\textbf{2. Despite using specialized logic, \am{the Edge TPU} operates far below its
theoretical maximum energy efficiency.} 
We use a similar approach to prior work~\cite{energy-roofline-model} to
 obtain a roofline for energy efficiency. 
Figure~\ref{fig:roofline-energy} (right) shows the
 roofline for the \am{Edge TPU}, along with the
efficiency achieved for each model.\revi{\footnote{\revi{Unlike} a throughput roofline, the energy roofline
 is a smooth curve because we cannot hide memory energy (as opposed
 to memory transfer time, which can be overlapped with computation time and results in the sharp knee
seen in throughput rooflines).}} We find that \revi{on average across
 all models, the Edge TPU achieves only 37.2\% of its maximum possible energy efficiency.} The energy efficiency is
 particularly low (33.8\% \revi{of the maximum}) for LSTM and Transducer models, but even the best CNN model
achieves only 50.7\% of the maximum efficiency.

\begin{figure}[h]
    \centering
        \centering
        \includegraphics[width=\linewidth]{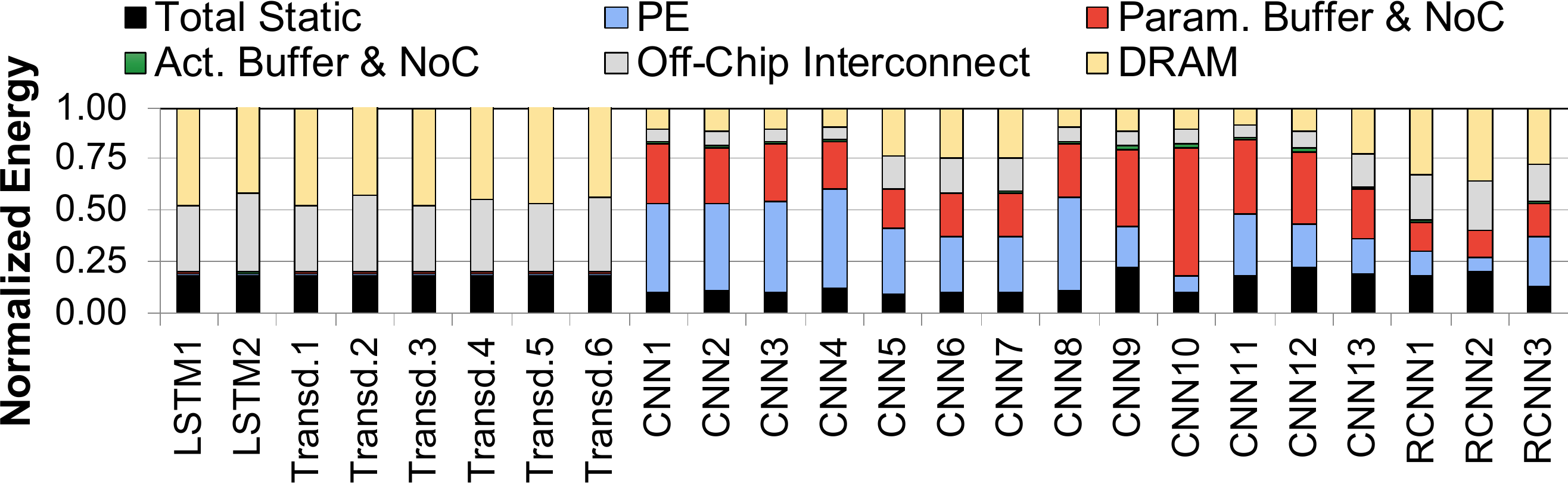}%
    \caption{Energy breakdown during inference execution.}
    \label{fig:energy-breakdown}
\end{figure}

\textbf{3. The accelerator's memory system design is neither effective nor efficient.} Figure~\ref{fig:energy-breakdown} shows the energy breakdown during inference
 execution across different models. We make three key observations from this figure. 
\revi{First, the on-chip buffers (the activation buffer and the parameter buffer) account for a significant portion of both static and dynamic energy across all models.
For example, for CNN models, 48.1\% of the total static energy and 36.5\% of the total dynamic energy is spent on accessing and storing parameters in the on-chip buffers.
This is due to the large size of both buffers in the Edge TPU.
Second, averaged across all models, the Edge TPU spends 50.3\% of its total energy on off-chip memory accesses (which includes the DRAM energy and the off-chip interconnect energy).
Third, for LSTMs and Transducers, the Edge TPU spends approximately three quarters of its total energy on DRAM accesses.
This is because while the buffers consume a significant amount of area
(79.4\% of the total area) in the Edge TPU, they are ineffective at
reducing off-chip memory accesses.
Despite the large buffers, only 11.9\% of the parameters for these models can fit into the buffer.
This is due to the parameter access patterns exhibited by LSTMs and Transducers: even if we ignore area constraints and increase the buffer capacity to 8x that of the Edge TPU, the buffer effectively caches only 46.5\% of the parameters (an increase of only 3.9x). Due to ineffective caching, the 8x buffer decreases latency by only 37.6\%, and energy consumption by only 40.3\%.
Overall, we conclude that the Edge TPU's overall memory system (which includes both on-chip buffers and off-chip memory) is highly inefficient, and results in significant energy consumption.}

\subsection{Layer-Level Study of Google Edge Models}
\label{sec:motiv:model}

To understand where \am{the Edge TPU's} \revi{shortcomings
(Section~\ref{sec:motiv:accelerator})}
come from, we analyze the models in significant detail, \revi{at the granularity of individual layers}.

\subsubsection{Analysis of LSTMs and Transducers}
\label{sec:motiv:model:lstm}

We identify three key properties of LSTMs and Transducers in our edge model analysis.

\begin{figure}[h]
    \centering
        \centering
        \includegraphics[width=\linewidth]{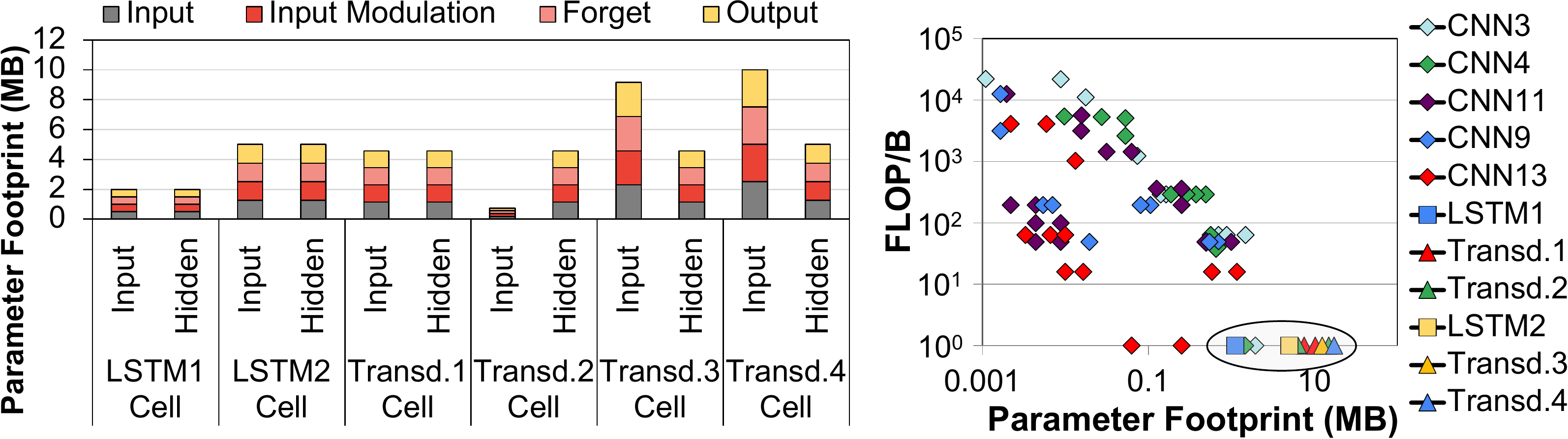}%
    \caption{Parameter footprint of $W_{x}$ and $W_{h}$ for different LSTM gates for LSTMs and Transducers (left)\revi{. L}ayer parameter footprint \revi{($W_{x}$ and $W_{h}$ combined)} vs.\ 
  \si{\flop\per\byte}, \revi{with LSTM/Transducer layers circled} (right).}
    \label{fig:footprint-lstm-cell}
    \label{fig:param-reuse}
\end{figure}

\textbf{1. Large parameter footprint.}
Each gate in an LSTM cell has an average of 2.1 million parameters, which
includes parameters for both input ($W_{x}$) and hidden ($W_{h}$) matrices
(as shown in Figure~\ref{fig:footprint-lstm-cell}, left).
The large parameter footprint of LSTM gates results in large footprints for
LSTM layers (up to 70 million parameters), and in turn, LSTM and Transducer models
that include such layers. Figure~\ref{fig:param-reuse} (right)
 shows the total footprint vs.\ the \si{\flop\per\byte} ratio (which indicates arithmetic intensity)
 across the layers of representative CNNs, LSTMs, and Transducers (the trend is the same across all models).
We observe from the figure \shepii{that} layers from LSTMs and Transducers have significantly larger footprints 
(with an average footprint of \SI{33.4}{\mega\byte}) than layers from CNNs.

\textbf{2. No data reuse and low computational 
complexity.} For these layers, the \si{\flop\per\byte} \revi{for parameters (which includes both $W_{x}$ and $W_{h}$)} is one
(Figure~\ref{fig:param-reuse}, right).
\revi{This is because the Edge TPU} fetches $W_{x}$ and $W_{h}$ for each LSTM gate from DRAM, accesses them once to
 perform the input and hidden MVMs \revi{for that gate}, and then does not touch the parameters again
 until the next LSTM cell computation, resulting in no reuse. \revi{In addition to the lack of reuse, LSTM and Transducer layers have much} lower computational complexity than \revi{CNN and RCNN layers},
with 67\% fewer MAC operations on average. 

\textbf{3. Intra- and inter-cell dependencies.} 
Two types of dependencies exist within LSTM layers, both of which
affect how the accelerator schedules \revi{LSTM gates}.
\revi{First, \emph{inter-cell dependencies} exist because
a cell with state $c_{t}$ needs the hidden vector from the
previous cell ($h_{t-1}$; see Section~\ref{sec:bkgd}) to perform the required MVMs for the four LSTM gates that make up the cell.
Second, \emph{intra-cell dependencies} exist because the hidden vector of the current cell ($h_{t}$) is computed using the outputs of the four LSTM gates in the cell, and thus cannot start until the cell state ($c_{t}$) is updated.} To respect these dependencies, the accelerator 
 schedules \revi{cell} computation in a
 sequential manner. \revi{Recall from Section~\ref{sec:bkgd} that each LSTM gate requires two MVMs: the input MVM and the hidden MVM. In order to compute these MVMs and respect dependencies, the Edge TPU} treats each gate as two
 fully-connected (FC) layers (corresponding to input MVM and hidden MVM),
 and runs the gates sequentially.

\revi{However, we find that this scheduling is inefficient. Specifically, the Edge TPU misses opportunities for parallelizing computation across the gates of a single cell. Instead, by treating the gates as multiple FC layers, the Edge TPU employs the same default layer serialization used for FC layers. While this ensures correctness for actual FC layers, this hurts LSTM performance, as the PEs to spend more time waiting for the MVMs to be completed due to the unnecessary serialization, which degrades PE utilization.
The lack of optimized support for LSTM cell computation results in significant missed opportunities to address LSTM performance and energy inefficiency.}

\subsubsection{Analysis of CNNs}
\label{sec:motiv:cnn}

\begin{figure}[h]
    \centering
        \centering
        \includegraphics[width=0.95\linewidth]{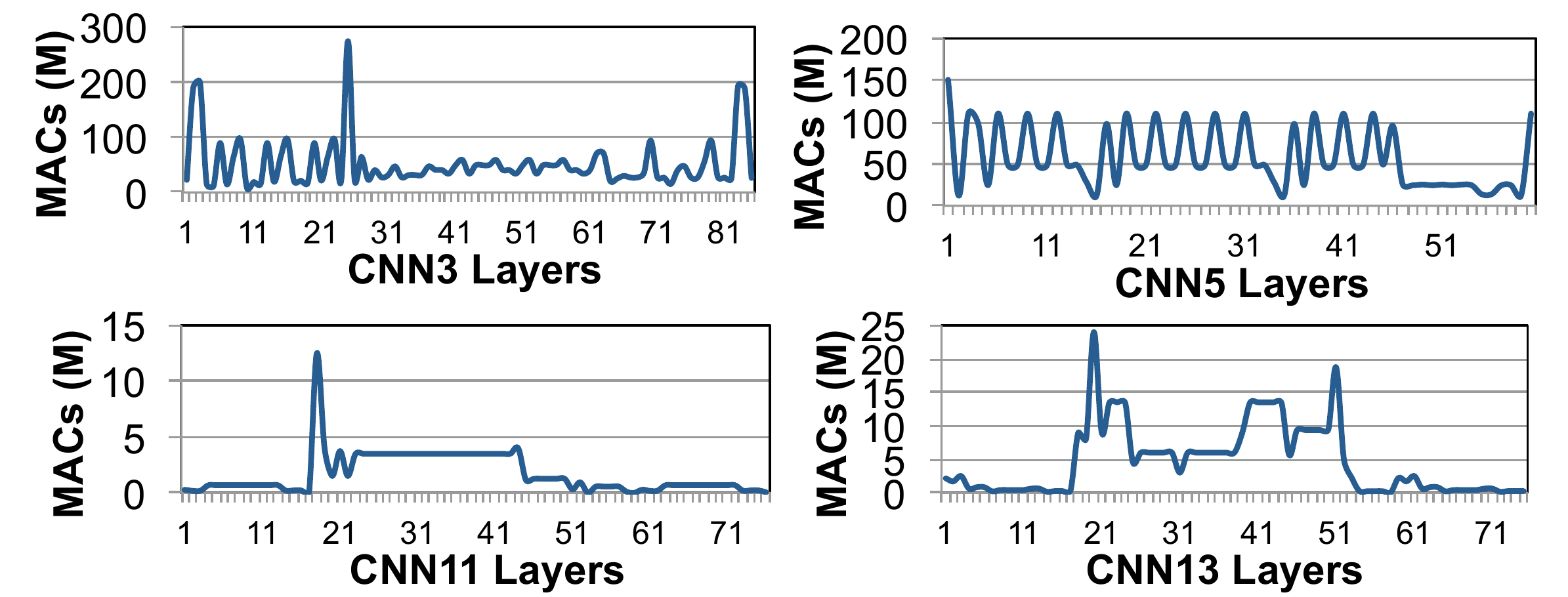}%
    \caption{\revi{MAC operation count in} different layers \revi{of} four CNN models.}
    \label{fig:cnn-mac}
\end{figure}

\begin{figure}[h]
    \centering
        \centering
        \includegraphics[width=0.95\linewidth]{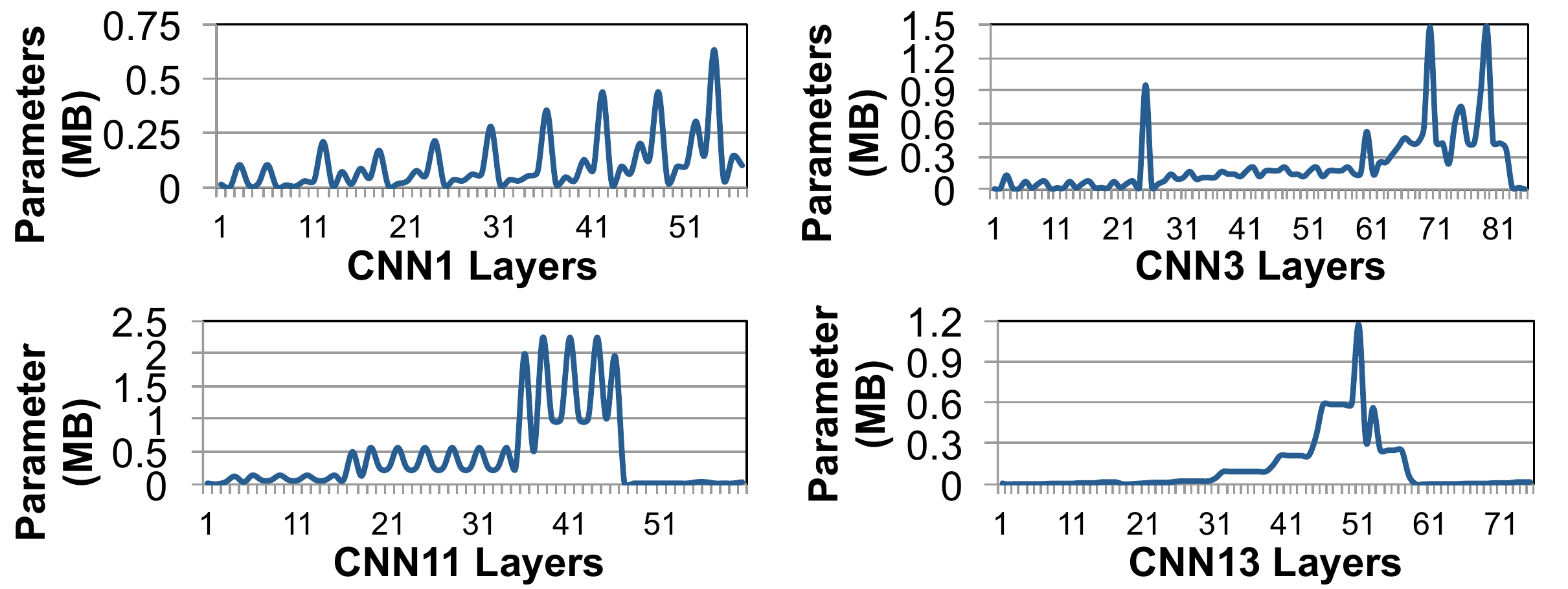}%
    \caption{Parameter footprint \revi{in} different layers \revi{of} four CNN models.}
    \label{fig:cnn-param}
\end{figure}

Our analysis of edge CNN models reveals two \revi{new} insights. First, unlike layers in traditional CNNs 
(e.g., AlexNet~\cite{alex-net}, VGG~\cite{simonyan2015very}), which tend to be relatively homogeneous, we find that the layers in
 edge CNN models exhibit significant heterogeneity in terms of type
 (e.g., depthwise, pointwise), shape, and size. 
\revi{This is often because these edge} models
 employ several decomposition techniques~\cite{mobilenet,squeezenet} to reduce the computational
 complexity and footprint of layers, in order to make them more friendly for the
 constrained edge devices. As an example of layer diversity,
Figures \ref{fig:cnn-mac} and \ref{fig:cnn-param} show the number of MAC operations and 
parameter footprint across different layers for four CNN models. 
We find that the MAC intensity and parameter footprint
 vary by a factor of 200x and 20x across different layers,
 \revi{indicating how significant the diversity is for important layer characteristics.}

Second, we find that layers exhibit significant variation in terms of data reuse patterns for
 parameters and for input/output activations. 
\revi{For example, a layer in a CNN can sometimes break down an
input into multiple filtered channels (e.g., breaking an image
down into red, green, and blue colors.
A pointwise layer operates on $K$ different channels,
and performs convolution using the same input activations on
each of these channels.
In comparison, a depthwise layer operates on only a single channel,
and is unable to reuse its input activations.} We also observe variation in data reuse across layers of the same type.
For example, initial/early standard convolution layers
in edge CNNs have a shallow input/output channel depth, large input activation
width/height, and very small kernels, resulting in very high
parameter reuse. In comparison, standard convolution
layers that are placed toward the end of the network have
a deep input and output channel depth, small input activation
width/height, and a large number of kernels, resulting in very low parameter reuse.  
This variation in reuse is illustrated in Figure~\ref{fig:param-reuse} (right),
 which shows that the \si{\flop\per\byte} ratio varies
across different layers for five representative CNN models
by a factor of 244x. 

\subsubsection{Analysis of RCNNs}

RCNNs include layers from both CNNs and LSTMs. As a result, individual layers from RCNN models
exhibit the same characteristics we \revi{describe} above for LSTM and CNN layers. 
We find that layers from RCNNs \revi{exhibit significantly} higher footprints and lower \si{\flop\per\byte}
ratios than \revi{CNN layers}, as \revi{an RCNN model includes} both LSTM and CNN layers. Due to the inclusion of both layer types, we observe significantly more variation across
RCNN layers characteristics as well. 

\subsubsection{Sources of \revi{Edge TPU} \revi{Shortcomings}}
\label{sec:motiv:discussion}

Using insights from our comprehensive model analysis
in Section~\ref{sec:motiv:model}, 
we can now \revi{provide more insight into} the sources of the \revi{Edge TPU}
shortcomings identified in Section~\ref{sec:motiv:accelerator}.

\paratitle{PE Underutilization} 
We identify three
 reasons why \revi{the actual throughput of the Edge TPU} falls significantly short of \revi{the peak}. First, while some layers have high parameter reuse
(e.g., pointwise layers, with a \SI{1200}{\flop\per\byte} ratio), 
other layers exhibit very low reuse (\SIrange{1}{64}{\flop\per\byte})
while \revi{at the same time} having large parameter footprints (\SIrange{0.5}{5}{\mega\byte}). \sg{Layers with low reuse yet large footprints for parameters
often leave PEs idle, as the parameters incur \revi{long-latency} cache misses to DRAM.}
The bandwidth of modern commercial DRAM
(e.g., \SI{32}{\giga\byte\per\second} for LPDDR4~\cite{lpddr4}) is two orders 
of magnitude below the \SI{2}{\tera\byte\per\second} bandwidth needed 
to sustain peak PE throughput when only one \si{\flop\per\byte} is performed
\sg{(as is the case for LSTM layers; \revi{see Section~\ref{sec:motiv:model:lstm}})}. \revi{Many layers end up with a 
similar memory bandwidth bottleneck during execution}.

Second, the \revi{Edge TPU} does \revi{\emph{not}} provide a custom dataflow optimized for each layer.
As we identified in Section~\ref{sec:motiv:model}, layers both across and within models
exhibit high variation in terms of data reuse patterns.
This variation necessitates the need for \emph{different} dataflows for different layers,
where each \revi{dataflow} exposes a different set of reuse opportunities for
parameters and activations.
However, \sg{state-of-the-art accelerators such as the Edge TPU employ} a \emph{single}
dataflow that is designed for high spatial/temporal reuse\revi{~\cite{meastro, edge-tpu, tetris, scnn, eyerissv2}}. The missed reuse opportunities in many of the model layers causes PEs
to needlessly wait on retrieving previously-accessed data that was not properly 
retained on-chip.

Third, the different shapes and inter-/intra-layer dependencies across 
different types of layers (e.g., LSTM cell, standard convolution, depthwise, 
pointwise, fully-connected)
makes it challenging to fully utilize a PE array with a fixed size,
which is the case in state-of-the-art accelerators \revi{(e.g., \cite{eyerissv2, edge-tpu, scnn, tetris})}.
To cater to these differences across layers, there is a need for both
better \revi{scheduling of MVM computation} (e.g., uncovering parallel computation opportunities
as we found in Section~\ref{sec:motiv:model:lstm})
and appropriately sizing the PE arrays \revi{based on the needs of specific layers}
in order to maintain efficient utilization.

\paratitle{Poor Energy Efficiency}
We find three major sources of energy \revi{efficiency in the Edge TPU}.
 First, the \revi{Edge TPU} incurs high static energy costs because (1)~it employs a large overprovisioned on-chip buffer, and (2)~it underutilizes PEs. Second, the on-chip buffers consume a high amount of dynamic energy, as we saw for CNN layers in Section~\ref{sec:motiv:accelerator}. Third, the \revi{Edge TPU} suffers from the high cost 
of off-chip parameter traffic. On-chip buffers
 fail to effectively cache parameters for many layers due to layer diversity, causing
 50.3\% of the total inference energy to be spent on
 off-chip parameter traffic \revi{(see Section~\ref{sec:motiv:accelerator})}.

\paratitle{Memory System Issues}
We uncover two \revi{large sources of memory system inefficiency}.
First, due to layer diversity, 
on-chip buffers are ineffective for a large fraction of layers. 
As we discuss in Section~\ref{sec:motiv:model:lstm},
LSTM gates have large parameter footprints and 
zero parameter reuse,
rendering the on-chip buffer useless for a majority of
LSTMs and Transducers, and for a significant fraction of RCNN layers
\revi{(i.e., models that incorporate LSTM layers)}. For CNN layers, we
find that those layers with low data reuse account for a significant portion
 of the entire model parameters (e.g., 64\% for CNN6). This means that the on-chip buffer fails
to cache a large portion of the parameters for CNN models. As a result, despite being \revi{\SI{4}{\mega\byte}} in size,
the on-chip \revi{parameter} buffer is effective only for a small fraction of layers,
which have an average parameter footprint of only \SI{0.21}{\mega\byte}.

Second, \revi{even layers with high data reuse incur significant costs when they access on-chip buffers.
This is because the on-chip parameter buffer is very large, even though the layers that benefit from caching in the buffer have small parameter footprints.
These layers exhibit high parameter reuse, and thus generate a large number of buffer accesses.
Unfortunately, because of the large size of the buffer, every access incurs a high dynamic energy cost.
As a result, this unneeded capacity for the high-data-reuse layers results in wasted energy for buffer accesses.}

\subsection{Key Takeaways}
\label{sec:motiv:takeaways}

\am{Our analysis provides three key insights:
(1)~there is significant variation in terms of layer characteristics \emph{across} and 
\emph{within} state-of-the-art Google edge models; 
(2)~the monolithic design of the Edge TPU
is the root cause of its shortcomings \revi{and the resulting large inefficiency;} and (3)~to achieve high utilization and energy efficiency, 
all key components of an edge accelerator (PE array, dataflow, on-chip memory,
off-chip memory bandwidth)
must be customized \revi{to different} layer characteristics. 
}


\section{\titleShort Framework}
\label{sec:proposal}
\label{sec:framework}

\titleShort is a new \revi{machine learning accelerator design} framework that \sg{harnesses \revi{inter- and intra-}layer variation across edge NN models for} high efficiency \revi{and high performance}.

\subsection{High-Level Overview}
\label{sec:framework:overview}

\revi{The key idea of \titleShort is to distribute} the layers from an NN model across
a collection of smaller hardware accelerators that are carefully specialized towards the
properties of different layer types. By specializing each accelerator to a subset of layers, \titleShort avoids 
\revi{the shortcomings of current monolithic edge ML accelerators}, 
resulting in \revi{a highly-efficient and high-performance} accelerator with a much smaller area. \titleShort consists of (1)~a collection of heterogeneous hardware accelerators;
and
(2)~a runtime scheduler that determines which accelerator each layer
in \revi{an NN} model should execute on, using a combination of \revi{NN} model and hardware
characteristics.
As we show in Section~\ref{sec:arch}, 
\revi{\titleShort designs typically need to employ only a small
number of accelerators, as
layers tend to group together into a small number of layer \emph{families}.}

\revi{We design \titleShort} as a framework that can support a wide range of
architectural implementations.  This allows \titleShort to 
\revi{(1)}~be optimized to
specific system needs, which is critical to keep resource utilization to a
minimum in resource-constrained edge devices; and \revi{(2)}~adapt easily to future types of NN models that we expect will arise in the future.
We discuss one example implementation of \titleShort in Section~\ref{sec:arch},
which caters to the Google edge \revi{NN} models that we analyze \revi{(Section~\ref{sec:motiv})}, to illustrate the effectiveness of our framework.

\subsection{\revi{\titleShort Runtime} Scheduler}
\label{sec:proposal:scheduler}
\label{sec:framework:scheduler}

The goal of {\titleShort}'s software runtime scheduler is to identify 
which accelerator each layer in an NN model should run on.
Each of the accelerators in \titleShort caters to 
\revi{one or more families of layers,
where these families share specific characteristics} (e.g., \revi{layer type}, footprint, data reuse, dependencies).
For a given \revi{\titleShort implementation}, the scheduler has two pieces of 
information \revi{(which can be maintained in a hardware driver)}:
(1)~the characteristics of each \revi{layer family}; and
(2)~which hardware accelerator is best suited for each \revi{family}.\revi{\footnote{%
\revi{This} information is generated
once during initial setup of a system, and can be modified with an updated
driver version to account for new \revi{families}.}}

\revi{\paratitle{Layer-to-Accelerator Mapping}} When an NN model runs on \titleShort, the scheduler generates a mapping 
between \revi{each NN layer and different} accelerators.
The scheduler uses the NN model (including a directed acyclic graph that 
represents communication across model layers) and the configuration information
in the driver to determine this mapping.\footnote{\label{fn:scheduler}\shep{In order to simplify the mapping process, 
\revi{our initial version of the \titleShort scheduler} does not
schedule multiple layers to run concurrently. 
\revi{Future versions can improve performance and efficiency by using a more sophisticated scheduler
that supports the concurrent mapping of multiple layers to multiple accelerators.}}} The mapping is generated in two phases. 

In \revi{Phase~I}, the scheduler iterates through each layer in the model, and 
identifies the ideal hardware accelerator for each layer \emph{in isolation}
(i.e., without considering communication overhead). The scheduler determines two properties for each layer: 
(1)~the cluster that the layer belongs to, and
(2)~the target accelerator for the layer. While \revi{the goal of Phase~I is} to maximize accelerator throughput and energy efficiency for each layer,
the resulting schedule may be sub-optimal for efficiency, because it does not
consider the overhead of transferring activations or communicating 
dependencies (e.g., $h_{t}$ in LSTM cells)
between different layers.
This can have a large impact on the overall \revi{system} performance and energy if
the amount of communication is large.

In \revi{Phase~II}, our scheduler accounts for the communication
overhead using a simple cost analysis algorithm, 
\revi{and assigns the destination accelerator for each layer}.
\revi{Phase~II iterates through each of the layers in a model sequentially,
and we describe the decisions made during Phase~II for
an arbitrary layer~$i$, which are performed after
layer~$i-1$'s destination accelerator (destination~$i-1$) has been assigned.
For layer~$i$, Phase~II determines whether the layer should be scheduled on
(1)~its ideal accelerator, as determined by Phase~I; or
(2)~destination~$i-1$.
Destination~$i-1$ is used whenever the costs of communicating
operands to the ideal accelerator outweigh the penalties of
executing layer~$i$ on destination~$i-1$, which is a sub-optimal
accelerator for the layer.\footnote{\revi{If destination~$i-1$ is the same
as the ideal accelerator for layer~$i$, the Phase~II analysis is
skipped for the layer, and the layer is assigned to its ideal accelerator.}}
There are two cases where Phase~II assigns layer~$i$ to its ideal accelerator.
First, if the number of MAC operations required for layer~$i$
is 2x higher (determined empirically) than the compute resources available
in destination~$i-1$, running the layer on destination~$i-1$
would incur increased performance and energy costs over using the
ideal accelerator.
Second, if the amount of parameter data that destination~$i-1$ 
would need to fetch to run layer~$i$ is greater than the
amount of output activation data that would have to be sent to
the ideal accelerator, \emph{and} the opportunities for reusing
the parameter data in destination~$i-1$ are low
(\si{\flop\per\byte} $<64$, determined empirically),
running the layer on destination~$i-1$ would incur
off-chip memory overheads (with few opportunities for amortizing
these overheads) compared to using the ideal accelerator.
In all other cases, Phase~II assigns layer~$i$ to destination~$i-1$.}

\shep{Mensa uses a heuristic-based approach that may not always achieve
the \revi{best} mapping decisions that a hypothetical oracle scheduler could produce.
However, our heuristic-based scheduler still achieves significant 
performance and energy improvements \revi{(Section~\ref{sec:eval})}, while being practical to implement in edge devices.
We leave the exploration of \revi{better} scheduling algorithms to future work.}

\revi{\paratitle{Execution and Communication}}
Once \revi{Phase~II of the scheduler} is complete, \titleShort begins model execution
using the generated \revi{layer-to-accelerator} mapping.
\revi{During execution, destination~$i$ needs to read
(1)~any unbuffered parameters (i.e., weights) from DRAM; and
(2)~input activations (i.e., input data) produced by layer~$i-1$, 
when layer~$i$ is run on a different accelerator than layer~$i-1$.
In order to simplify communication between accelerators,
\titleShort accelerators transfer activations to another accelerator through DRAM,
avoiding the need to keep on-chip data coherent across
accelerators (or, when some of the \titleShort accelerators are
placed near memory, to keep on-chip and near-data accelerators coherent~\cite{conda,LazyPIM}).}


\section{\revi{\exampleDesign:} \titleShort for Google Edge Models}
\label{sec:arch}

We now discuss \revi{\exampleDesign,}
\sg{an example \titleShort design optimized} for our Google edge \revi{NN}
models.
We start by identifying \revi{layer families} in these models
(Section~\ref{sec:arch:clusters}).
Using the unique characteristics for each \revi{family}, we determine which
characteristics have the greatest impact on accelerator design, and use
that to guide the number of accelerators that we need \revi{for Google edge NN models}
(Section~\ref{sec:arch:hw}).

\subsection{Identifying \revi{Layer Families}}
\label{sec:proposal:analysis}
\label{sec:arch:clusters}
\label{sec:arch:families}

\begin{figure}[h]
    \centering
        \centering
        \includegraphics[width=\linewidth]{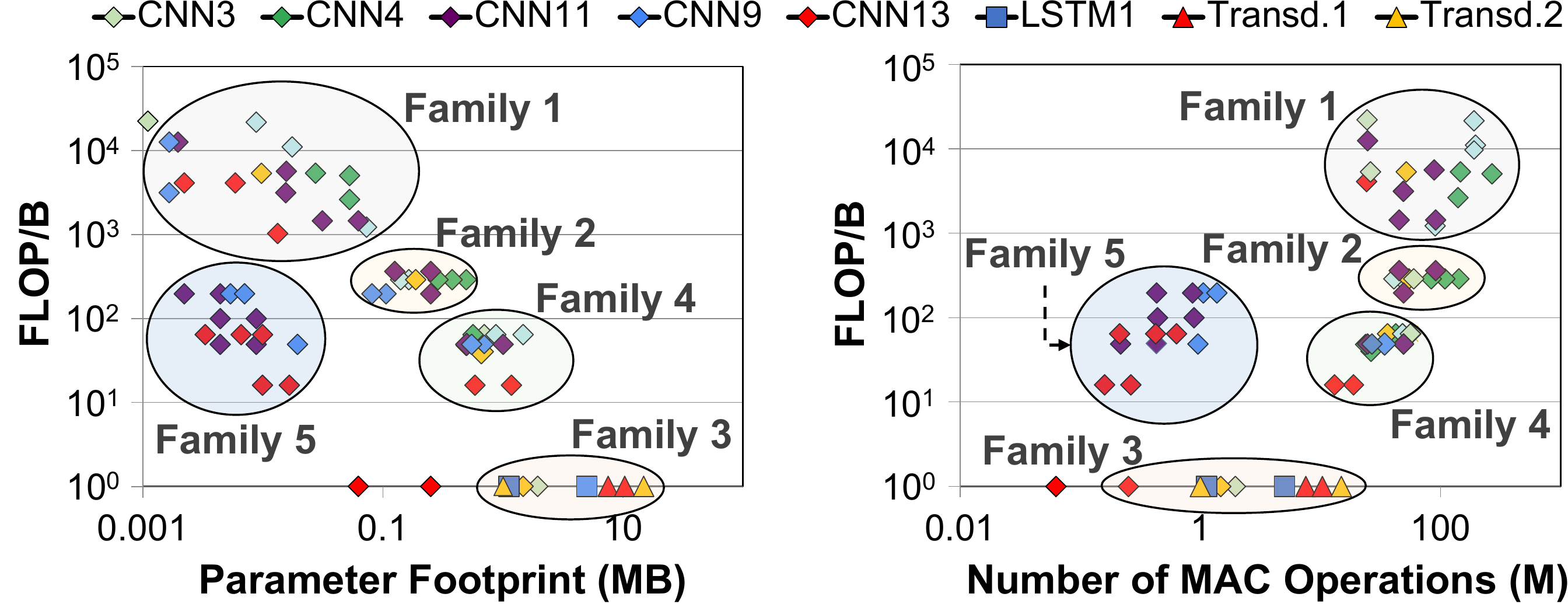}%
    \caption{Parameter footprint vs. \revi{parameter reuse} (left) and number of MAC operations vs. \revi{parameter reuse} (right) across \revi{example} layers.}
    \label{fig:layer-analysis}
\end{figure}

We revisit the \revi{NN edge} models that we analyze in Section~\ref{sec:motiv}.
For each layer, we study the correlation between different characteristics. 
\revi{These characteristics include 
(1)~parameter reuse (\si{\flop\per\byte}), 
(2)~parameter footprint (\si{\mega\byte}), and
(3)~MAC intensity (i.e., defined by the number of MAC operations).} \revi{Figure~\ref{fig:layer-analysis} shows
how parameter reuse
(\si{\flop\per\byte}) correlates \revi{with} the parameter footprint (Figure~\ref{fig:layer-analysis}, left) and the 
number of MAC operations (Figure~\ref{fig:layer-analysis}, right)
for a representative set of layers from 
five CNNs, two LSTMs, and two Transducers.}\footnote{\revi{We show a representative set of layers in Figure~\ref{fig:layer-analysis} to improve the figure's clarity. We do not include layers from RCNNs as they consist of CNN and LSTM layers.}}
Based on all of the layer characteristics that we analyze,
we observe across all layers from all models (not just the representative
layers or correlations plotted) that 97\% of the layers group into one of
five \revi{layer families}.

\paratitle{\revi{Family}~1} 
Layers in this \revi{family} have 
(a)~a very small parameter footprint (\SIrange{1}{100}{\kilo\byte}),
(b)~a very high \si{\flop\per\byte} ratio \revi{for parameters} (780--20K), and
(c)~high MAC intensity (30M--200M). 
These layers exhibit high activation 
footprints and \revi{high activation} data reuse as well.
A majority of layers in this \revi{family} are standard \revi{convolutional} 
layers with shallow input/output channels and large input activation width/height.
We find that these layers are mostly found among early layers \revi{in CNNs and RCNNs}, and typically achieve high PE utilization (on average 82\%) \revi{on the Google Edge TPU because of their high MAC intensity and low memory use}. 

\paratitle{\revi{Family}~2}
Layers in this \revi{family} have 
(a)~a small parameter footprint (\SIrange{100}{500}{\kilo\byte};
12x higher on average \revi{than} \revi{Family}~1),
(b)~a moderate \si{\flop\per\byte} ratio \revi{for parameters} (81--400;
up to 10x lower than \revi{Family}~1), and 
(c)~high MAC intensity (20M--100M). \revi{These layers} exhibit high activation footprints
and activation data reuse as well. Many of the layers belong to pointwise layers, which have high parameter reuse
due to convolving 1x$K$ filters (\revi{where $K$ is the input channel depth, i.e., the number of channels}) with input \revi{activations} across different channels.
Other layers in the \revi{family} include standard convolution layers commonly
found in the middle of CNN networks, with 
deeper input/output channels and smaller activation width/height than the 
convolution layers in \revi{Family}~1, 
We find that \revi{Family}~2 layers have lower PE utilization (64\%)
than layers in \revi{Family}~1 \revi{on the Google Edge TPU, because the lower parameter reuse reduces opportunities to amortize the off-chip memory access overheads}.

\paratitle{\revi{Family}~3} 
Layers in this \revi{family} have 
(a)~a very large parameter footprint (\SIrange{0.9}{18}{\mega\byte}),
(b)~minimal \revi{\si{\flop\per\byte} ratio for parameters,} and (c)~low MAC intensity (0.1M--10M).
These layers \revi{also} exhibit small activation footprints \revi{but high} activation reuse. 
The majority of these layers are from LSTM gates in LSTMs and Transducers,
or are fully-connected layers from CNNs. These layers have very low PE utilization (0.3\% on average) \revi{on the Google Edge TPU, as there are not enough MAC operations to hide the significant off-chip memory bottlenecks incurred while retrieving parameters}.

\paratitle{\revi{Family}~4}
Layers in this \revi{family} have 
(a)~a \revi{relatively large} parameter footprint (\SIrange{0.5}{2.5}{\mega\byte}),
(b)~low-to-moderate \si{\flop\per\byte} ratio for parameters (25--64), and
(c)~moderate MAC intensity (5M--25M). \revi{These} layers exhibit small activation footprints \revi{but high} activation reuse. A large portion of layers in this category are standard \revi{convolutional} layers with
deep input/output channels and input activation width/height,
along with a large number of kernels. \revi{Family}~4 layers \revi{have a low PE utilization (32\% on average) on the Google Edge TPU, as the large parameter footprint and relatively low parameter reuse generate significant off-chip memory traffic, and the moderate MAC intensity hides only some of the memory access bottlenecks.}

\paratitle{\revi{Family}~5} 
Layers in this \revi{family} have 
(a)~a very small parameter footprint (\SIrange{1}{100}{\kilo\byte}),
(b)~a moderate \si{\flop\per\byte} ratio for parameters (49--600), and
(c)~low MAC intensity (0.5M--5M). \revi{These} layers \revi{exhibit high} activation footprints but have almost zero activation data
reuse. 
\revi{Many of the layers in Family~5 are depthwise convolution layers.
Such layers have only one channel, and thus do not reuse activations, and typically have only a small number of filters (where each filter consists of one or more kernels) that are applied to all slices of the inputs, resulting in high parameter reuse.} \revi{Family}~5 layers achieve a low average PE utilization of 21\% \revi{on the Google Edge TPU, as the increased parameter reuse compared to Family~4 is offset by the reduced MAC intensity, and Family~5 layers still have limited opportunities to hide the memory access bottlenecks}.

\subsection{Hardware Design Principles \revi{and Decisions}}
\label{sec:proposal:implication}
\label{sec:arch:hw}

As we study the distinguishing characteristics of each \sg{of \revi{the} five \revi{layer families}}, we find that
some characteristics have a strong influence on the hardware design,
while others do not necessitate significant changes to the
hardware.
We discuss two insights that drive our hardware design decisions.

First, we find that \revi{significantly different values of MAC intensity and parameter footprint/reuse lead to greatly different hardware to maximize efficiency}, as they impact a number of
key accelerator design parameters (e.g., PE array size,
on-chip buffer size, memory bandwidth considerations).
Looking at \revi{the} five \revi{layer families}, we identify that
(a)~layers in \revi{Families} 1 and 2 share a high MAC intensity, 
\revi{small} parameter footprint, and moderate-to-high parameter reuse; while
(b)~layers in \revi{Families} 3 and 4 share a low MAC intensity,
large parameter footprint, and low parameter reuse.
This means that we need \emph{at least} two different accelerator
designs:
one that caters to the compute-centric behavior of \revi{Families}~1/2 \revi{(see Section~\ref{sec:arch:comp-x})},
and one that caters to the data-centric behavior of \revi{Families}~3/4 \revi{(see Section~\ref{sec:arch:mem-x:dataflow})}.
Given our resource-constrained \revi{edge} environment, we look to see if layers in
\revi{Family}~5, which have a low MAC intensity (similar to \revi{Families} 3 and 4) but
a \revi{relatively} small parameter footprint (similar to \revi{Families} 1 and 2),
can benefit from one of these two approaches.
We find that the low MAC intensity, along with the \revi{low} parameter reuse
by many \revi{Family}~5 layers, allow the layers to benefit from many of the
non-compute-centric optimizations that benefit \revi{Families} 3 and 4, so we \revi{study them collectively as we design the data-centric accelerators}.

Second, 
\sg{a key distinguishing factor between different accelerator designs
is the accelerator dataflow, as it dictates which reuse opportunities in layers are exploited,
and thus strongly impacts PE utilization and energy efficiency.} \revi{One p}rior work~\cite{meastro} analyzes the large dataflow design space,
and discusses four types of data reuse:
\emph{spatial multicasting} (reading a parameter once, and spatially \revi{distributing}
it as an input to multiple PEs \revi{at the same time}),
\emph{temporal multicasting} (replicating a parameter in a small local buffer,
and delivering the parameter as multiple inputs at different times to the same PE),
\emph{spatial reduction} (accumulating activations from multiple PEs 
at the same time using multiple compute units), and
\emph{temporal reduction} (accumulating multiple activations generated at 
different times using a single accumulator/buffer).
The chosen dataflow
directly affects how the memory system and on-chip network of an accelerator
should be designed.
Thus, we need different dataflows for \revi{layer families} with
significantly different \revi{parameter and activation} reuse patterns.

\subsubsection{\revi{Determining the Number of Accelerators Needed}}

Both of our compute-centric \revi{layer families (Families 1 and 2)} \sg{benefit from a similar}
dataflow, which exposes reuse opportunities for both parameters and
activations.
Between the compute-centric optimizations and the shared dataflow affinity,
we determine that we can use a single accelerator 
(\emph{\accelA}\revi{; Section~\ref{sec:arch:accelA}}) to efficiently execute layers from both \revi{Family~1 and Family}~2.

Across our three data-centric \revi{layer families (Families 3, 4, and 5)}, we find that
\sg{layers from both \revi{Families} 4 and 5} benefit from a dataflow that exposes
reuse opportunities for parameters
but not for activations,
and can \revi{use a single} accelerator
(\emph{\accelC}\revi{; Section~\ref{sec:arch:accelC}}). \revi{Family}~3 layers exhibit different data reuse characteristics, \revi{and they benefit} from a dataflow that provides temporal reduction
 opportunities for activations.
As a result, \sg{they} \revi{benefit from} a separate accelerator
(\emph{\accelB}; \revi{Section~\ref{sec:arch:accelB}}).
  
\subsubsection{\revi{Template-Based Design Approach}}

We employ a \emph{template-based} design approach \revi{for the compute- and data-centric accelerators}:
while we design each accelerator \revi{so that it is} based on \revi{layer families'} characteristics, we \revi{use} the
same generic tiled architecture \revi{for each accelerator} as the baseline \revi{edge TPU}. 
We do this to ease the integration of our hardware into a real system:
from the perspective of compilers and programs, each of our accelerators
appears to be just a different instance (with a different configuration) of
\revi{a single} baseline accelerator \revi{(the Edge TPU)}.
This is a \emph{critical design decision}\revi{:} it allows us to
 deploy and run models using existing highly-optimized design/compile toolchains
(e.g., \revi{the Google} Edge TPU compiler~\cite{edge-tpu-compiler}) 
seamlessly on all of \titleShort's accelerators, \revi{but it also} \revi{limits} the degree of customization (and, thus, efficiency) \revi{each of} our
accelerator designs can achieve.
While \revi{the} \titleShort~\revi{framework easily allows the design of} accelerators that do not employ this
tiled architecture, we make this design choice to
\revi{reduce the burden on the software stack (e.g., more complex compilers, multiple libraries
for programmers)}.

\shep{The three accelerators in \revi{\exampleDesign}
are designed to be independent:
we customize each accelerator's dataflow, on-chip memory (i.e., buffers), and access
to off-chip memory for the layers that the accelerator targets,
and \revi{carefully} provision the number of PEs and design the interconnect
to support the chosen dataflow and memory access patterns.
To simplify the design and increase modularity, we do \revi{\emph{not}} share
any resources between accelerators.}

\subsection{\revi{\accelA: Compute-Centric} Accelerator Design}
\label{sec:arch:accelA}
\label{sec:arch:comp-x}

\begin{figure}[h]
    \centering
    \includegraphics[width=\columnwidth]{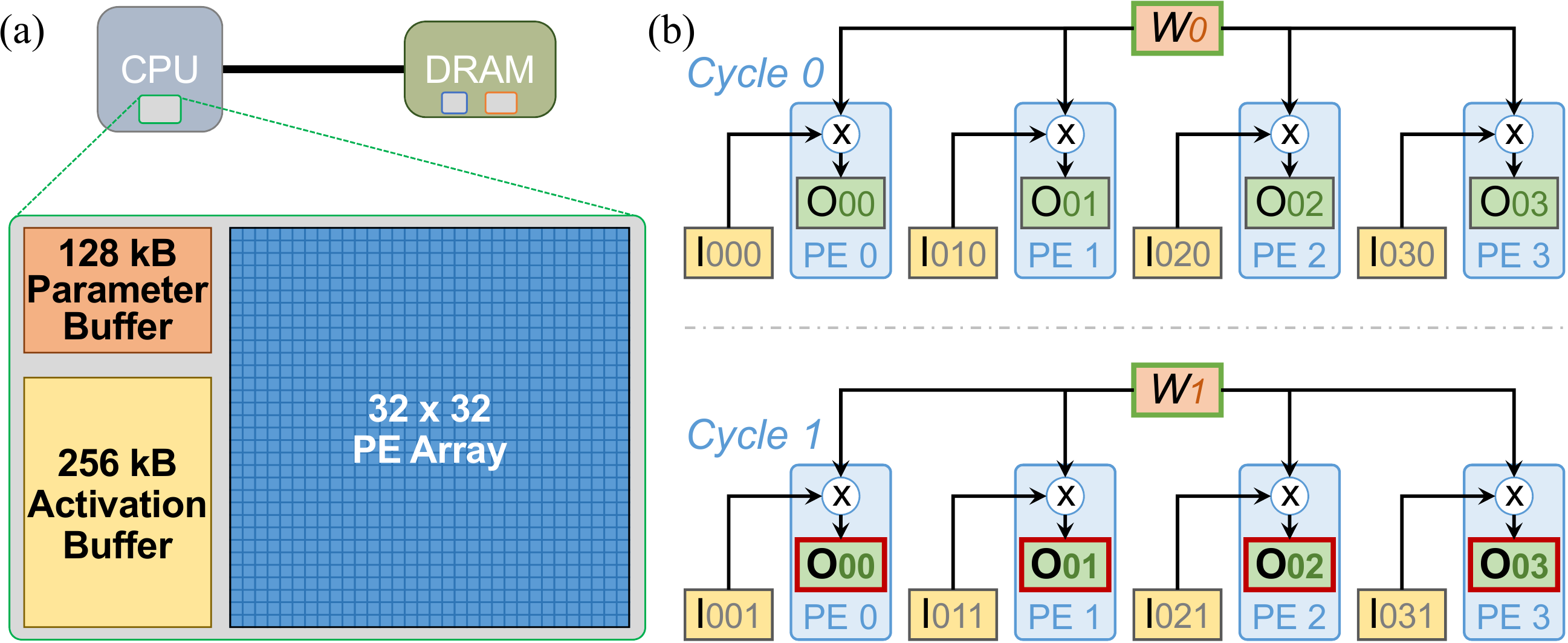}
    \caption{(a)~\accelA design; (b)~\accelA dataflow for a pointwise layer.}
    \label{fig:accelA}
\end{figure}

\revi{\accelA caters to layers in Families 1 and 2, which are compute-centric.
We establish two requirements for the design of \accelA.
First, the design should exploit opportunities available across layers in Families 1 and 2 for the temporal reduction of output activations.
The temporal reduction mitigates the impact of the large output activation footprint, by reducing the off-chip memory bandwidth requirements and providing an opportunity to reduce the buffer capacity needed for activations.
Second, the design should \emph{avoid} spatial reduction for output activations.
Spatial reduction generates partial output sums in each PE, and then gathers all of the partial sums by sending them across the on-chip network to a single PE.
Given the large footprint of output activations, this partial sum traffic often saturates the limited bandwidth of the on-chip network, which can leave the PEs underutilized and lead to significant performance and energy penalties.
Based on these two requirements, we design \accelA as shown in Figure~\ref{fig:accelA}a.}

\paratitle{Dataflow}
\label{sec:arch:comp-x:dataflow}
\revi{Figure~\ref{fig:accelA}b shows the dataflow that we design for \accelA.
In the figure, we show how the dataflow works for an example pointwise layer in a CNN,
where the layer has $K$~channels.
For each channel, every element in the matrix is multiplied by a single parameter (i.e., a weight representing a filter is applied point by point).
In the output activation matrix $O$, each element $O_{ij}$ (where $i$ is the row, and $j$ is the column) is computed as the sum of $I_{ijk} \times W_{k}$ over every channel~$k$, where $I$ is a three-dimensional matrix of input activations and $W$ is a vector of weights.}

\revi{As shown in Figure~\ref{fig:accelA}b, \accelA's dataflow reduces memory traffic by enabling two types of reuse.
First, the dataflow uses temporal reduction for each output activation element $O_{ij}$ \emph{without} using spatial reduction, by instead having a single PE accumulate the entire sum of the element across multiple cycles in the PE's private register file. 
We indicate temporal reuse (i.e., temporal multicasting) in the dataflow using a red border and bold text in the figure.
Second, the dataflow uses spatial multicasting for each parameter $W_{k}$, by ensuring that all of the PEs are working on the same channel $k$ in the same cycle.
We indicate spatial multicasting using a green border and italic text in the figure.}

\paratitle{PE Array}
\label{sec:arch:comp-x:pe}
\revi{We size the PE array in \accelA based on two attributes.
First, there should be enough PEs to accommodate the high MAC intensity exhibited by layers in Families 1 and 2.
Second, all PEs should ideally operate on the same parameter in a single cycle, to minimize parameter bandwidth.
To account for both attributes, and to ensure a good balance between PE utilization, inference latency, and energy consumption, we profile the performance of Family~1/2 layers on different PE sizes, and empirically choose a 32x32 PE array,
which lets \accelA achieve a \SI{2}{\tera\flop\per\second} peak throughput.}

\paratitle{Memory System}
\label{sec:arch:comp-x:mem}
\revi{Figure~\ref{fig:accelA}a shows the on-chip buffers used in \accelA.
These buffers are significantly smaller than those in the Google Edge TPU.
We reduce the size of the activation buffer from \SI{2}{\mega\byte} in the Edge TPU to \SI{256}{\kilo\byte} in \accelA, because \accelA's dataflow exploits temporal reduction for the output activations using the internal PE registers, and no longer needs to store the large footprint of the output activations in the activation buffer.
We reduce the size of the parameter buffer from \SI{4}{\mega\byte} in the Edge TPU to \SI{128}{\kilo\byte} in \accelA, because layers in Families 1 and 2 have small parameter footprints.
Given the low off-chip memory bandwidth requirements of \accelA, we keep the accelerator on the CPU die.}

\subsection{\revi{\accelB: LSTM-Centric} Accelerator Design}
\label{sec:arch:accelB}
\label{sec:arch:mem-x}
\label{sec:arch:mem1-x}

\begin{figure}[h]
    \centering
    \includegraphics[width=\columnwidth]{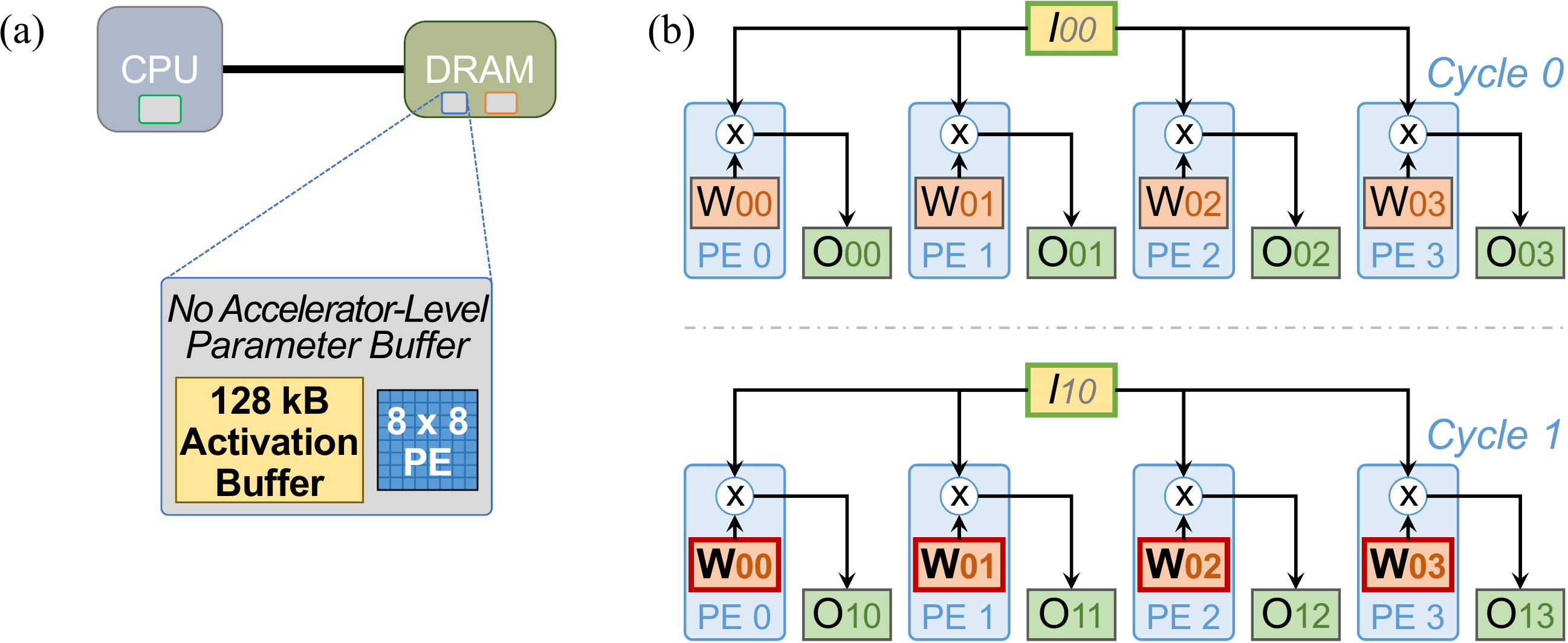}
    \caption{(a)~\accelB design; (b)~\accelB dataflow.}
    \label{fig:accelB}
\end{figure}

\revi{\accelB caters to layers in Family~3, which are data-centric \emph{and} predominantly consist of LSTM layers.
We establish two requirements for the design of \accelB.
First, the design should exploit output activation reuse opportunities in Family~3 layers.
Second, the design should reduce the off-chip memory bandwidth required by parameters. These layers have a very large parameter footprint, and parameters that are cached in the parameter buffer of the Google Edge TPU are evicted before they can be reused, forcing every parameter access to go to main memory.
Based on these two requirements, we design \accelB as shown in Figure~\ref{fig:accelB}.}

\paratitle{Dataflow}
\label{sec:arch:mem-x:dataflow}
\label{sec:arch:mem1-x:dataflow}
\revi{Figure~\ref{fig:accelB}b shows the dataflow that we design for \accelB.
Recall from Section~\ref{sec:bkgd} that an LSTM layer has multiple cells, each containing four gates. Each gate performs two MVMs (an input MVM, which multiplies the input parameter matrix $W_{x}$ with the input vector $x_{t}$; and a hidden MVM, which multiplies the hidden parameter matrix $W_{x}$ with the hidden input vector $h_{t-1}$, across a sequence of samples over time ($t$).
To generalize the figure for both MVMs, we show a single MVM between a parameter matrix $W$ (where each element $W_{ij}$ is located at row~$i$, column~$j$) and an input activation vector set $I$ (where each element $I_{ti}$ is the located at row~$i$ in the input activation vector for time~$t$), where $I \times W = O$ ($O$ is the output vector set for a particular gate).}

\revi{As shown in Figure~\ref{fig:accelB}, \accelB's dataflow reduces memory traffic by enabling two types of reuse.
First, the dataflow temporally reuses a weight $W_{ij}$.
We observe that in an LSTM layer, instead of iterating one cell at a time (where we compute the input MVMs and hidden MVMs for the four gates), we can compute the input MVMs for all $C$~cells in the layer back-to-back, and then compute the hidden MVMs for all four gates.
This allows the dataflow to fetch each element of $W$ only once per layer (as opposed to fetching each element $4TC$ times, once for every sample $t$ for each of the four gates, over $C$~cells). To enable the temporal reuse of the weight, the PE stores the weight in one of its private registers,
and stores $C$~partial sums (which are accumulated over time to generate $C$~outputs for each cell in the layer).
We indicate temporal reuse (i.e, temporal multicasting) in the dataflow using a red border and bold text in the figure.
Second, the dataflow uses spatial multicasting for each input activation $I_{ti}$, as the same activation is multiplied across all columns of $W$ for a given row~$i$. We indicate spatial multicasting using a green border and italic text in the figure.}

\paratitle{PE Array}
\label{sec:arch:mem-x:pe}
\label{sec:arch:mem1-x:pe}
Because \revi{Family~3} layers \revi{have low} MAC intensity,
and mainly perform MVM,
we design a much smaller PE array for \accelB than that in the
\revi{Google Edge TPU}.
\revi{We analyze} the inference latency across a range of
PE array sizes, and \revi{empirically choose an 8x8 array size to balance} latency, utilization, and energy.
\revi{This allows \accelB to achieve a \SI{128}{\giga\flop\per\second} peak throughput.}

\paratitle{Memory System}
\label{sec:arch:mem-x:mem}
\label{sec:arch:mem1-x:mem}
\revi{We decide to place \accelB \emph{inside memory}~\cite{pim-survey, pim-book, ghose.pim.bookchapter18}, to accommodate the significant off-chip memory bandwidth requirements of Family~3 layers.
Modern 3D-stacked memories such as High-Bandwidth Memory~\cite{hbm} and the Hybrid Memory Cube~\cite{hmcspec2} include logic layers that have access to the high memory bandwidth available within a 3D-stacked memory chip.
By placing \accelB in the logic layer of a 3D-stacked memory chip, we can provide Family~3 layers with much higher bandwidth than the external memory bandwidth.}

Given that parameters and activations from
\revi{Family~3 layers} exhibit different characteristics, we 
customize separate on-chip buffers for each data type, \revi{as shown in Figure~\ref{fig:accelB}a}. 
For parameters, we use only one level of memory hierarchy
(\SI{512}{\byte} \revi{of private registers} per PE), \revi{eliminate the parameter buffer,} and stream parameters directly
from DRAM.
\revi{The per-PE registers provide enough space to cache the temporally-multicasted parameters, and there are no other reuse opportunities that an accelerator-level parameter buffer could exploit (because the parameters have a very large footprint).} For activations, thanks to the small
activation footprint of layers in \revi{Family}~3, we use a \SI{128}{\kilo\byte} buffer.

\subsection{\accelC Accelerator Design}
\label{sec:arch:accelC}
\label{sec:arch:mem2-x}

\begin{figure}[h]
    \centering
    \includegraphics[width=\columnwidth]{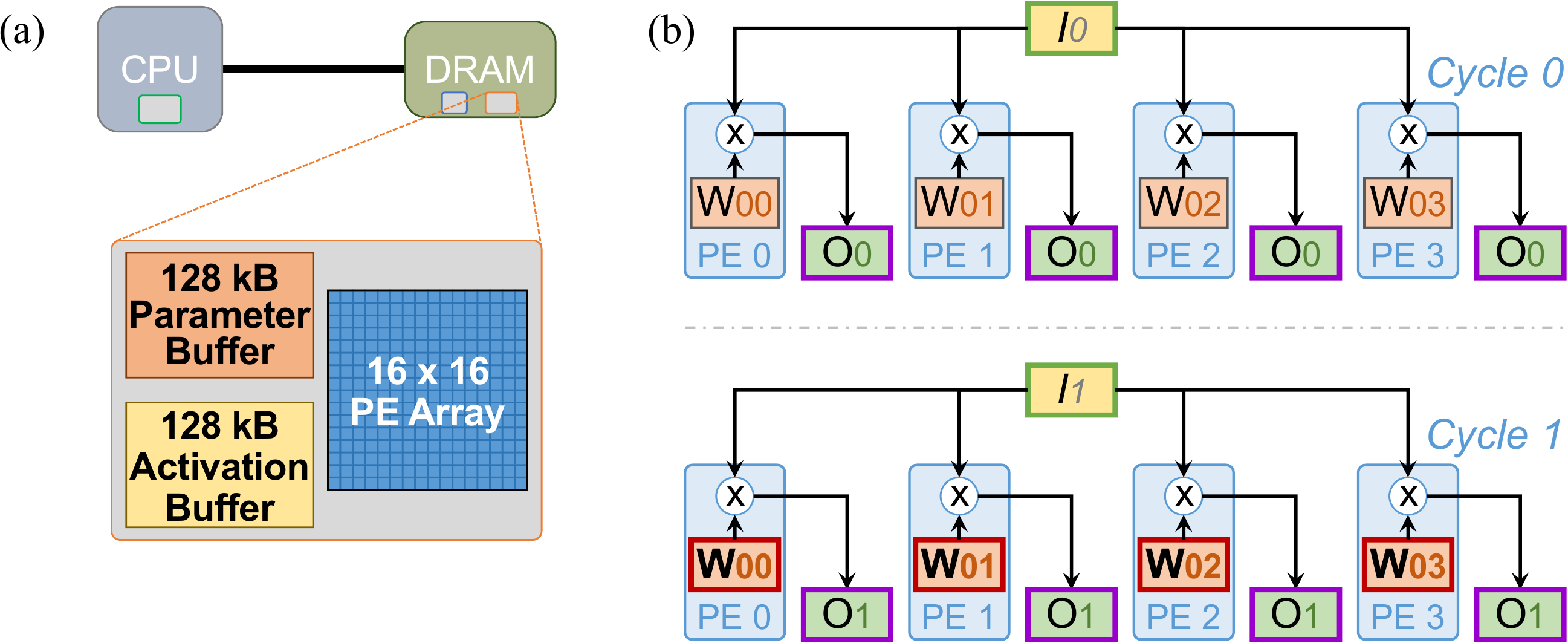}
    \caption{(a)~\accelC design; (b)~\accelC dataflow.}
    \label{fig:accelC}
\end{figure}

\revi{\accelC caters to layers in Families 4 and 5, which primarily consist of non-LSTM data-centric layers.
We establish two requirements for the design of \accelC.
First, the design should exploit temporal reuse opportunities for parameters in layers from Families 4 and 5.
Second, the design should provide high off-chip memory bandwidth, as the parameter footprint is high for many (but not all) of the layers.
Based on these two requirements, we design \accelC as shown in Figure~\ref{fig:accelC}.}

\paratitle{Dataflow}
\label{sec:arch:mem2-x:dataflow}
\revi{Figure~\ref{fig:accelB}c shows the dataflow that we design for \accelB.
We illustrate the dataflow using a generic MVM, where an input activation vector~$I$ is multiplied by a parameter matrix~$W$ to generate an output activation vector~$O$.
As shown in the figure, \accelC's dataflow reduces memory traffic by enabling two types of reuse.
First, the dataflow temporally reuses a parameter $W_{ij}$.
\accelC uses the same temporal reuse approach as \accelB, where a parameter is stored in a PE register and reused (i.e., temporally multicast) over multiple cycles to reduce the number of times the parameter is fetched from memory.
By increasing the reuse of the parameter, the dataflow effectively hides the off-chip memory access latency by overlapping it completely with PE computation.
We indicate temporal reuse in the dataflow using a red border and bold text in the figure.
Second, the dataflow uses spatial multiicasting for each input activation $I_{i}$. We indicate spatial multicasting using a green border and italic text.
In order to enable spatial multicasting for layers in Families 4 and 5, the dataflow has all PEs collectively compute a single output activation, by having each PE compute a partial sum (indicated in the figure with a purple border), and then using the on-chip interconnect to gather the partial sums and produce the final output activation.}

\paratitle{PE Array}
\label{sec:arch:mem2-x:pe}
While \revi{layers in Families 4 and 5 have low} MAC intensity, they perform
more MAC operations on average than \revi{Family}~3 layers. 
Our analysis of PE array sizes shows that 
equipping \accelC with an array smaller than 16x16 increases the latency,
so we \revi{empirically} select 16x16 for the array size. This allows \accelC to achieve a peak throughput of
\SI{512}{\giga\flop\per\second}. 

\paratitle{Memory System}
\label{sec:arch:mem2-x:memory}
Similar to \accelB, we decide to place \accelC inside the logic layer of 3D-stacked memory.
\revi{Doing so enables high memory} bandwidth for the large parameter footprints of \revi{Family}~4 layers. We use separate shared buffers for each data type, \revi{as shown in Figure~\ref{fig:accelC}a}. Given the small activation footprints, we use
a small \SI{128}{\kilo\byte} buffer for them (a 16x reduction compared to the \revi{Google Edge TPU}). \sg{Thanks to the temporal parameter reuse that \revi{\accelC's} dataflow enables, we can reduce the 
 parameter buffer size \revi{to \SI{128}{\kilo\byte} (a 32x reduction compared to the Google Edge TPU)}.}

\subsection{\shep{Data Transformations \& Communication}}
\label{sec:arch:comm}

\shep{The three accelerators \revi{in \exampleDesign}
are designed to replace the core of the Edge TPU (i.e., the monolithic PE array and interconnect).
Our Mensa implementation maintains all of the other hardware support
from the overall Edge TPU architecture.
Notably, Mensa supports the same data transform 
operations (e.g., the \emph{im2col} operation for convolution, which takes the 
current window of pixels being processed in an image and 
converts the data layout into a matrix column) 
that the Edge TPU currently performs.
Such data transform operations are typically handled during
local data exchanges between the PEs.}

\shep{In \revi{\exampleDesign}, the accelerators communicate with each other only \revi{in-between} layer execution,
as layers do not execute concurrently \revi{(see Footnote~\ref{fn:scheduler})}, and each layer executes completely in a single accelerator.
We observe that \revi{Google} edge models typically communicate between accelerators \revi{only} 4--5 times during execution. 
Three of our models (CNN5, CNN6, CNN7) \revi{communicate significantly more frequently than average, as they} include a large number of \emph{skip connections} (i.e.,
layer~$i$ takes in data that was output by layer~$i-j$, where $j>1$). Mensa schedules the layers from these models across multiple accelerators, \revi{and}
the target accelerator has to fetch information produced by early layers (e.g., output feature maps) 
from either off-chip memory or from another accelerator's on-chip buffer (if the data has not yet been evicted). 
Mensa's scheduler coordinates this \revi{inter-layer and inter-accelerator} communication, and our evaluations take the extra \revi{communication traffic} into account.}


\section{Experimental Methodology}
\label{sec:methodology}

\paratitleattop{}
\paratitle{Models}
\sg{The \revi{24} Google edge NN models that we analyze} are used in several
 Google mobile applications/products, such as image classification, object detection,
 semantic segmentation, automatic speech recognition, and image captioning. The models are
 specifically developed for edge devices using TensorFlow Lite~\cite{tensorflow-lite}, and
 are fully 8-bit quantized using \sg{quantization-aware training}~\cite{quant-aware-training}.
 The models are then compiled using the \revi{Google} Edge TPU compiler~\cite{edge-tpu-compiler}. \shepii{We expect to see similar results for popular publicly-available models such as 
MobileNet~\cite{mobilenet} and ResNet~\cite{he.cvpr2016},
which share similarities with some of our edge models.}
 
\paratitle{Energy Analysis}
We build our energy model based on prior works~\cite{google-pim, tetris, conda}, which sums up the total energy 
\shep{(including both static and dynamic energy)} consumed by
the accelerator, DRAM, off-chip and on-chip interconnects, and all on-chip buffers. We use CACTI-P 6.5~\cite{CACTI} with a \SI{22}{\nano\meter} process technology
 to estimate on-chip buffer energy. We assume that each 8-bit MAC unit consumes \SI{0.2}{\pico\joule\per\bit}. We model
 the DRAM energy as the energy consumed per bit for LPDDR4~\cite{lpddr4}, based on models from prior works~\cite{tetris, google-pim}.

\paratitle{Performance Analysis \revi{\& Simulation}}
We use an in-house simulator \sg{to \revi{faithfully} model all major components of} the \revi{Google} Edge TPU, including the PE array, memory
 system, on-chip network, and dataflow. We heavily modify the simulator to implement our three proposed accelerators
and the software runtime of \titleShort. 
We develop an analytical cost model to determine the performance of each of our proposed dataflows, and integrate
the dataflow performance numbers into our simulator's performance model. We use CACTI-P 6.5~\cite{CACTI} \sg{to determine the on-chip buffer latencies for each proposed accelerator. 
Similar to prior works~\cite{google-pim, tetris, mondrian, conda}, 
the accelerators in the logic layer of 3D-stacked memory have access to
the \SI{256}{\giga\byte\per\second} internal bandwidth of High-Bandwidth Memory (HBM)~\cite{hbm},
which is 8x the \revi{external memory} bandwidth to accelerators that sit
outside of memory.} \shep{In our evaluation, we assume that both the Edge TPU baseline and \titleShort
have access to \SI{2}{\giga\byte} of HBM DRAM.}


\section{Evaluation}
\label{sec:eval}

We evaluate \revi{inference} energy \revi{(Section~\ref{sec:eval:energy})}, \revi{hardware utilization and throughput (Section~\ref{sec:eval:performance})}, \revi{and inference latency (Section~\ref{sec:eval:latency})} \revi{of} \am{four} configurations:
(1)~\emph{Baseline}, the \revi{Google} Edge TPU;
(2)~\emph{Base+HB}, a hypothetical version of Baseline with 8x \revi{the memory} bandwidth (\SI{256}{\giga\byte\per\second}); 
\am{(3)~\emph{\revi{Eyeriss~v2}}, a state-of-the-art edge accelerator~\cite{eyerissv2} that uses reconfigurable interconnects to \revi{partially} address CNN model heterogeneity; and}
(4)~\revi{\emph{\exampleDesign} (from Section~\ref{sec:arch})} with all three proposed accelerators (\accelA, \accelB, \accelC).
To improve figure clarity, we show individual model results
for only a few representative models of each model type.
Our average results are reported across \emph{all} 24 Google edge \revi{NN} models.

\subsection{Energy Analysis}
\label{sec:eval:energy}

\begin{figure*}[t]
    \centering
        \includegraphics[width=\linewidth]{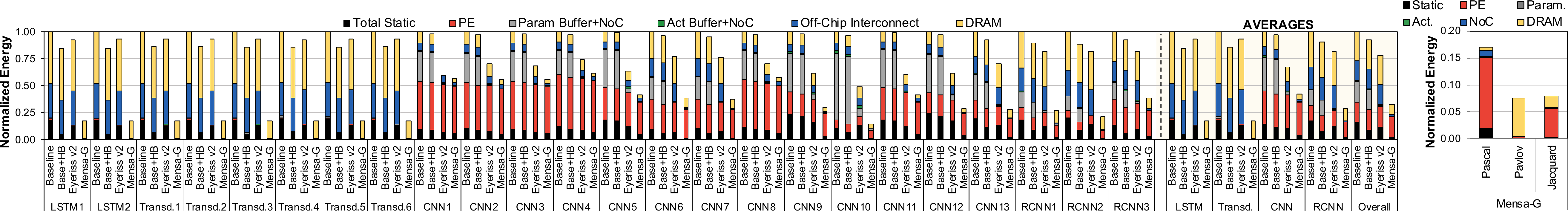}%
    \caption{Inference energy across different models (left) and energy breakdown across our three proposed accelerators (right), normalized to Baseline.}
    \label{fig:eval:energy-breakdown}
   \vspace{-10pt}
\end{figure*}

Figure~\ref{fig:eval:energy-breakdown} (left) shows the total inference energy \revi{consumed by the four systems we evaluate} across
different NN models. We make three observations from the figure.

First, providing high \revi{memory} bandwidth to Baseline \revi{(Base+HB)} results in only a small
reduction \revi{in} energy \revi{(7.5\% on average)}.
This is because Base+HB still incurs \revi{a high energy cost due to}
(1)~on-chip buffers that are overprovisioned for many layers, and
(2)~off-chip traffic to DRAM. Base+HB benefits LSTMs and Transducers the most (14.2\% energy reduction),
as the higher \revi{memory} bandwidth significantly reduces the inference latency of these models,
which in turn lowers static energy.

\sg{Second, \revi{Eyeriss~v2}
 suffers from significant energy inefficiency for LSTMs and Transducers. While \revi{Eyeriss~v2} lowers static energy
  compared to Baseline, due to its use of a much smaller PE array (384 vs.\ 4096) and on-chip buffers (\SI{192}{\kilo\byte} vs.\ \SI{4}{\mega\byte}), it still
 incurs the high energy costs of \revi{large} off-chip parameter traffic to DRAM. Averaged across all LSTM and Transducer models,
 \revi{Eyeriss~v2} reduces energy by only 6.4\% over Baseline. For CNN models, \revi{Eyeriss~v2} reduces energy by 36.2\% over Baseline, as its
smaller on-chip buffer significantly reduces dynamic energy consumption.}

Third, \revi{\exampleDesign} significantly reduces energy across all models.
The reduction primarily comes from three sources.
(1)~\revi{\exampleDesign} lowers the energy spent on on-chip and off-chip parameter traffic
by 15.3x \revi{over Baseline}, by scheduling \revi{each} layer on the \revi{accelerator} with the most appropriate
dataflow for \revi{that} layer.
LSTMs and Transducers benefit the most, as their energy \am{in both Base+HB and \revi{Eyeriss~v2}}
is dominated by off-chip parameter traffic,
which \accelB and \accelC drastically \revi{reduce by being placed} inside memory.
(2)~\revi{\exampleDesign} reduces the dynamic energy of the on-chip buffer and
network (NoC) by 49.8x \am{and 6.2x over Base+HB and \revi{Eyeriss~v2}}, by avoiding overprovisioning
and catering to specialized dataflows.
This is most beneficial for CNN and RCNN models.
(3)~\revi{\exampleDesign} reduces static energy \am{by 3.6x and 5.6x over Base+HB and \revi{Eyeriss~v2}}, thanks to
using significantly smaller PE arrays that avoid underutilization,
significantly smaller on-chip buffers, and dataflows that reduce inference latency.

\am{\revi{Eyeriss~v2} falls significantly short of Mensa's energy efficiency for three reasons. First, while
 \revi{Eyeriss~v2}'s flexible NoC can provide a high data rate to the PE array, its \revi{\emph{fixed}} dataflow cannot efficiently expose
 reuse opportunities across different layers (e.g., \revi{Family} 4 and 5 layers that have very large parameter
footprints and low data reuse). 
\sg{Second, \revi{Eyeriss~v2} has much higher static energy consumption, 
as its inference latency is significantly larger for many compute-intensive CNN layers
(as its PE array is much smaller than \revi{\accelA's PE array}).} Third, some CNN layers have a large parameter footprint and very low data reuse, which
generates a large amount of off-chip parameter traffic \sg{in \revi{Eyeriss~v2}}.} Overall, \revi{\exampleDesign} reduces total energy by \am{66.0\%/50.6\%, and improves energy efficiency
(\si{\tera\flop\per\joule}) by 3.0x/2.4x, compared to Baseline/\revi{Eyeriss~v2}}.

Figure~\ref{fig:eval:energy-breakdown} (right) shows the breakdown of
energy usage across \revi{the} three \revi{\exampleDesign} accelerators.
\revi{Compute-centric} \accelA consumes the most energy of the three, with its
consumption dominated by the PE array (since the layers that run on
\accelA perform a large number of MAC operations).
\revi{LSTM-centric} \accelB's energy usage is dominated by DRAM accesses, as its layers
have large footprints and no data reuse.
For \revi{data-centric} \accelC, the majority of energy is used by a combination of DRAM accesses
and the PE array, but the usage is lower than \accelB DRAM accesses or
the \accelA PE array due to the inherent layer properties (smaller footprints,
lower MAC intensity).

\subsection{\revi{Utilization and Throughput Analyses}}
\label{sec:eval:performance}

\begin{figure}[h]
    \centering
        \centering
        \includegraphics[width=\linewidth]{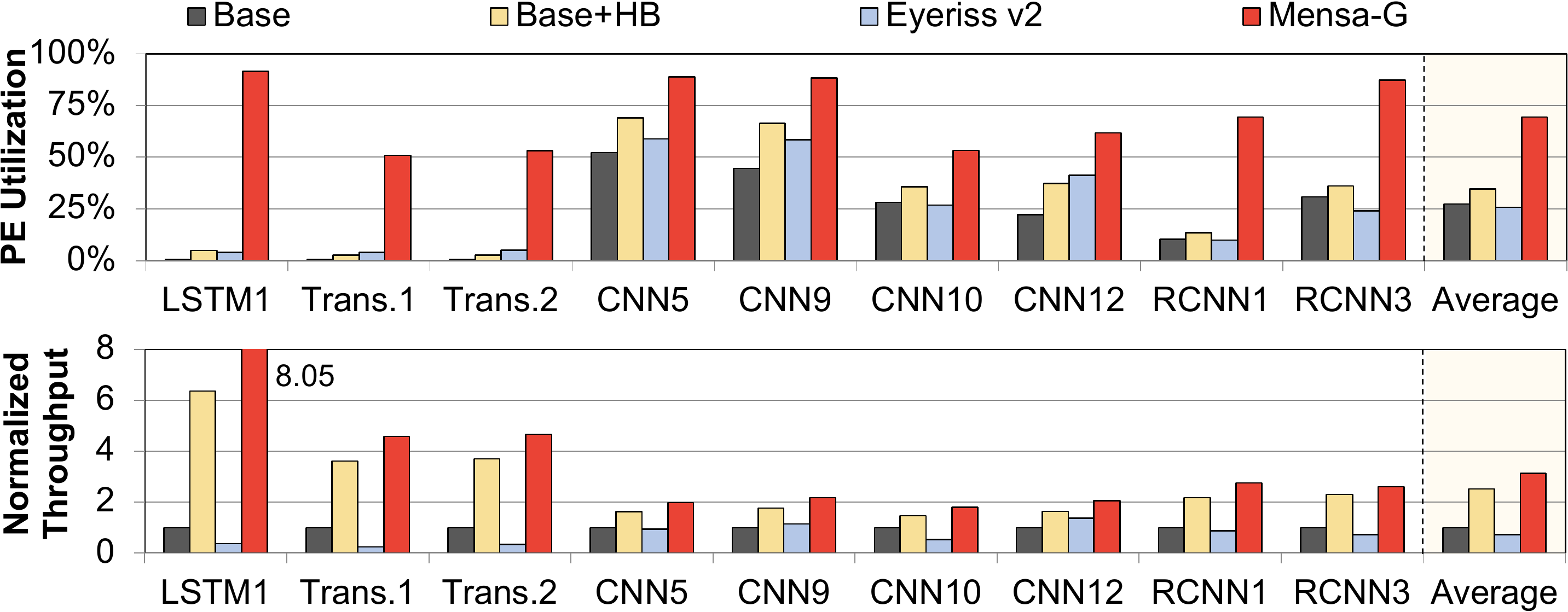}%
    \caption{\revi{PE} utilization \revi{(top)} and \revi{Baseline-normalized} throughput \revi{(bottom)} across different models.}
    \label{fig:eval:utilization}
\end{figure}

Figure~\ref{fig:eval:utilization} shows the
\revi{raw PE} utilization (\revi{top}) and \revi{Baseline-normalized} throughput \revi{(bottom)}
for our \am{four} configurations. \revi{\exampleDesign}'s utilization is calculated by computing
the average utilization across its three accelerators (\accelA, \accelB, and \accelC). 
We make three observations from the figure.

First, Baseline suffers from low PE array utilization (on average 27.3\%). The higher \revi{memory} bandwidth in Base+HB \revi{increases the} average \revi{PE} utilization to 34.0\%,
and improves throughput by 2.5x.
The largest \revi{throughput} improvements \revi{with Base+HB} are for LSTMs and Transducers (4.5x on average
vs.\ 1.3x for CNNs), thanks to their low
\si{\flop\per\byte} ratio and large footprints.
In contrast, some CNN models (e.g., CNN10) see only modest improvements (11.7\%)
with Base+HB, as their layers have high reuse and small footprints.
Overall, Base+HB still has very low utilization, as many
layers (those from \revi{Families} 3, 4, and 5) do not need the large number of
PEs in the accelerator. 

\sg{Second, \revi{Eyeriss~v2} \emph{reduces} performance significantly over Baseline
for several models.  \revi{Eyeriss~v2}'s flexible interconnect and much 
 smaller PE array allow it to achieve slightly higher PE utilization than Baseline for layers with very low data
 reuse. 
 However, this higher utilization is offset by significantly higher inference latencies.
For compute-intensive layers in \revi{Families} 1 and 2, the smaller PE array size hurts layer throughput.
For data-\revi{intensive} layers in \revi{Families} \revi{3,} 4 and 5, \revi{Eyeriss~v2} cannot customize its dataflow
to expose reuse opportunities, and thus is \revi{greatly} hurt
by the high off-chip traffic \revi{(particularly for LSTMs and Transducers)}.
Overall, we find that \revi{Eyeriss~v2}'s overall throughput is actually \emph{lower}
than Baseline for most of our models.}

\am{Third, \revi{\exampleDesign} \revi{significantly increases} both average utilization
(\revi{by} 2.5x/2.0x/2.6x) and throughput
(\revi{by} 3.1x/1.3x/4.3x) over Baseline/Base+HB/\revi{Eyeriss~v2}.} The large utilization improvements are a result of
(1)~properly-provisioned PE arrays for each layer,
(2)~customized dataflows that exploit reuse and opportunities for parallelization, and
(3)~the movement of large-footprint layer computation into \revi{3D-stacked} memory
(\revi{which eliminates} off-chip traffic for their DRAM requests).
We note that \revi{\exampleDesign}'s throughput improvements over Base+HB
are smaller than its utilization improvements\revi{. Even though Base+HB achieves poor energy efficiency and underutilization for layers with poor reuse and large footprints, it} is reasonably effective at reducing the inference latency
\revi{for} such layers. \revi{\exampleDesign} benefits all NN model types, but the largest improvements are for
LSTMs and Transducers, with average utilization/throughput improvements of
82.0x/5.7x over Baseline.
The improvement is lower for CNNs and RCNNs (2.23x/1.8x over Baseline),
because they make \revi{better} use of Baseline's large PE arrays, and
have smaller footprints that lessen the impact of off-chip DRAM accesses.
\revi{For a few CNNs (CNN10--CNN13) that use a large number of depthwise layers (which belong to \revi{Family}~5),
\exampleDesign's PE utilization is somewhat lower than desired (44.7\%) due to \revi{the} depthwise layers. These CNNs have significantly lower data reuse than other Family~4/5 layers, which in turn, makes these layers run less optimally with \accelC's dataflow. However, \exampleDesign still improves PE utilization for depthwise layers
by 65.2\% over Baseline, as a result of \accelC's specialization.}

\subsection{\revi{Latency Analysis}}
\label{sec:eval:latency}

\begin{figure}[h]
        \centering
        \includegraphics[width=\linewidth]{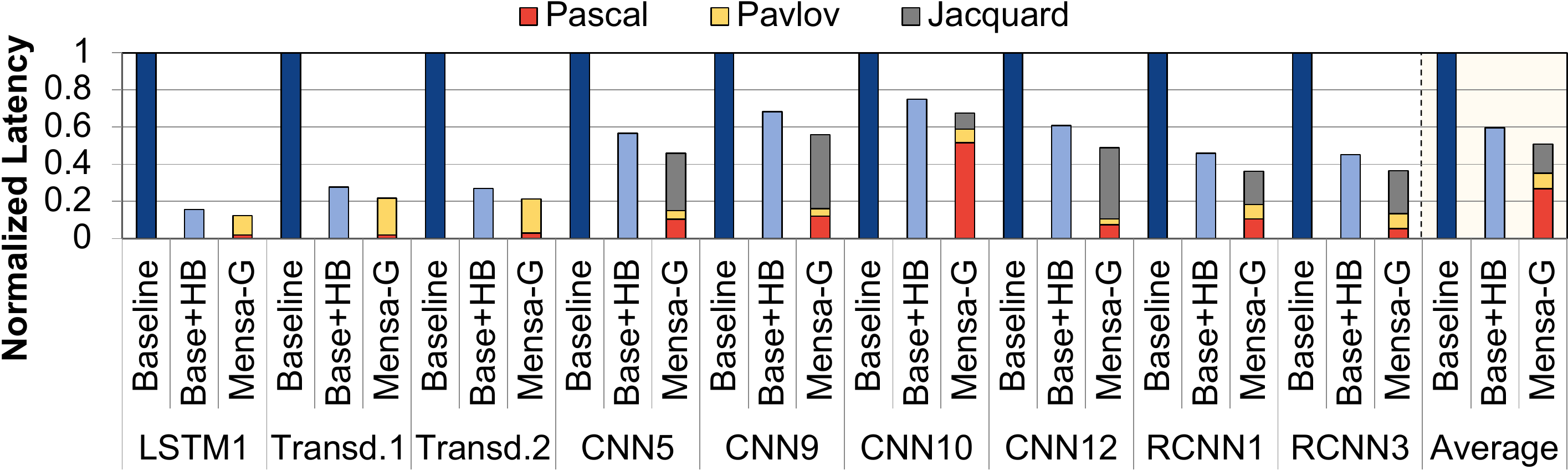}%
    \caption{Inference latency, normalized to Baseline.}
    \label{fig:eval:latency}
\end{figure}

Figure~\ref{fig:eval:latency} shows the \revi{Baseline-normalized} inference latency, and \revi{its breakdown across the three} \revi{\exampleDesign accelerators} (\accelA, \accelB, \accelC).
We find that \revi{\exampleDesign reduces inference latency over} Baseline and Base+HB on average by 1.96x and 1.17x.
LSTMs and Transducers see a significant latency reduction with \revi{\exampleDesign}
(5.4x/1.26x vs.\ Baseline/Base+HB) because most of their layers
run on \accelB and benefit from an optimized dataflow and processing-in-memory
(which provides not only higher bandwidth, but also lower latency for DRAM accesses).
CNNs and RCNNs benefit from the heterogeneity of our accelerators,
making use of all three of them to reduce latency by 1.64x/1.16x over Baseline/Base+HB.

\section{\revi{A Case for \titleShort Beyond the Edge TPU}}
\label{sec:takeaways}

\paratitle{\revi{Accounting for Diversity in Other State-of-the-Art Accelerators}}
\revi{Our evaluation of \exampleDesign shows that \titleShort can significantly improve performance and energy by replacing the core of the Google Edge TPU (i.e., the PE array and interconnect) with multiple accelerators.}
\shep{While our characterization and analysis focus on the \revi{Google} Edge TPU,
the \revi{resulting} insights can also be applied to several state-of-the-art
accelerators that are based on systolic arrays
(e.g., Eyeriss~v2~\cite{eyerissv2}, \revi{MAERI}~\cite{maeri}, HDA~\cite{kwon.hpca2021}).
The fundamental design of these accelerators, like the Edge TPU, is centered
around an array of PEs that are connected using on-chip networks 
to each other and to on-chip buffers\revi{. During inference, the accelerators} orchestrate a specific dataflow across the interconnected PEs,
buffers, and off-chip memory.
Systolic-array-based accelerators deployed for edge inference
will need to be designed to efficiently accommodate the very diverse
behavior that we uncover across and within edge \revi{neural network} models.
We note that the edge models that we evaluate include
(1)~several \revi{models that are widely used in the community}, and 
(2)~emerging models \revi{(e.g., LSTMs, Transducers)} that are expected to be of \revi{high} importance
in the near future.
As a result, our characterization can be helpful in informing
the diversity that future edge inference accelerators should capture.}

\paratitle{\revi{Adopting Multi-Accelerator Frameworks}}
\shep{\revi{We} believe that there is a strong case for
adopting \revi{the \titleShort framework} for future accelerator designs.
Given the tight area and energy resource budgets available for \revi{machine learning inference},
\revi{it has become critical to maximize} the efficiency of the accelerator
hardware, even at the expense of some additional design effort.
Our example implementation of \titleShort for Google edge models
demonstrates \revi{that} we can \emph{decrease} energy costs compared to
a monolithic general-purpose accelerator \revi{by tailoring multiple accelerators according to the characteristics of different NN models}.
We believe that \titleShort can work \revi{together with} other existing
accelerator designs beyond the Edge TPU,
where \titleShort manages multiple instances of the accelerator
with each instance customized (at design time and/or runtime)
and properly provisioned for a subset of models.}

\paratitle{\revi{Providing Flexibility to Incorporate New NN Model Types}}
\shep{As our study exposes, the emergence of new model types \revi{(e.g., LSTMs, Transducers)}
can often demand different resource trade-offs to achieve \revi{high} efficiency.
With the popularity of machine learning today, we expect that new models
will emerge that do not run optimally on any of our three proposed
accelerators.
With \revi{the \titleShort framework, which uses a software-level scheduler
to coordinate the mapping of layers to accelerators}, we can significantly reduce
design and deployment costs for new accelerators that cater
to these new models. This is because \titleShort's scheduler can easily incorporate
the attributes of new accelerators and identify which layers
to map to them, without requiring significant changes to the runtime software.}

\section{Related Work}
\label{sec:related}

\sg{To our knowledge, this is the first work to
(1)~examine the bottlenecks of a state-of-the-art commercial
Google Edge TPU when executing state-of-the-art Google edge \revi{neural network}
models;
(2)~quantify the significant \revi{intra- and inter-}layer variation that exists in state-of-the-art
edge NN models;
(3)~identify that layers can be clustered together based on a number of
shared \revi{execution} characteristics;
(4)~propose a new framework for heterogeneous ML inference acceleration (\revi{called} \titleShort),
with both on-chip and near-data accelerators; and
(5)~provide \revi{and evaluate the performance and energy benefits of} an example heterogeneous accelerator design for
Google edge NN models.}

\paratitle{Studies on a Single \revi{NN} Model Type}
Many prior works~\cite{e-pur, eyeriss, tetris, eyerissv2, shidiannao, serving-rnn, 
masr, scaledeep, rana, wax, cnn-resource, recnmp, cnvlutin, fused-layer, 
diannao, flexflow, scnn, simba} \revi{examine} a specific model type
(predominantly CNNs). None of these works
perform an analysis across different classes of edge \revi{NN} models (e.g., Transducers, LSTMs, RCNNs).
In fact, many \revi{studies} (e.g., \cite{eyeriss, 
tetris, shidiannao, simba, scnn}) analyze traditional models (e.g., AlexNet\revi{~\cite{alex-net}}, VGG\revi{~\cite{simonyan2015very}}),
and their \revi{acceleration} proposals are not tailored toward state-of-the-art edge \revi{NN} models (which we show are different)
or resource-limited edge devices. These proposals \revi{customize} accelerators
toward a particular model type (e.g., CNNs~\cite{eyeriss, tetris, rana, cnn-resource, scnn},
 LSTMs~\cite{e-pur,serving-rnn}), and \revi{thus} they are not optimized to serve multiple model types. 
As a result, they all suffer from the \revi{accelerator shortcomings} we discuss \revi{in detail} in Section~\ref{sec:motiv}.

\paratitle{Studies \revi{on NN Heterogeneity}}
Some CNN-focused works observe diversity
across CNN layers~\cite{meastro, scaledeep, eyerissv2, maeri, neurosurgeon, simba}. \shep{\revi{MAERI}~\cite{maeri} proposes an accelerator with a
reconfigurable on-chip network that connects various building blocks together. 
\revi{MAERI}'s network can be configured to
support different dataflows between PEs.
Eyeriss~v2~\cite{eyerissv2} includes a reconfigurable
on-chip network as well, but uses a single
dataflow that can support a wide range of CNN models. Despite the network reconfigurability, both
\revi{MAERI} and Eyeriss~v2
\sg{(1)~cannot \revi{customize} a number of essential
design parameters (e.g., off-chip memory bandwidth, on-chip memory) \revi{to different layers},
(2)~require frequent online reconfiguration to \revi{adapt} to \revi{intra- and inter-}layer
variation, and
(3)~make it difficult to co-design the dataflow with key components such as the memory system.}} \sg{We compare \revi{\exampleDesign to} Eyeriss~v2  in Section~\ref{sec:eval} \revi{and show that \exampleDesign provides significantly higher performance and energy efficiency}.}

\shep{ScaleDeep~\cite{scaledeep}
 proposes customized processing tiles to address diversity in CNN layers. 
While the idea shares
some similarities with \titleShort (exploiting heterogeneity in hardware),
there are two key differences.
First, while \titleShort targets inference using \revi{a diverse set of} edge \revi{NN} models \revi{(including LSTMs, Transducers)},
ScaleDeep targets cloud-based training of \revi{mainly} traditional CNN-based models.
The resource trade-offs are significantly different between the two targets:
edge \revi{NN} inference optimizes for tight area and energy constraints,
while cloud-based training optimizes for energy efficiency \revi{and performance} at \revi{very} large scale.
Second, due to \revi{its} focus on CNNs,
\revi{ScaleDeep} does not address the \revi{extensive} diversity that exists between CNN layers
and layers in other model types (e.g., LSTM layers).} Neurosurgeon~\cite{neurosurgeon} \revi{examines} at both vision and speech models. However, 
their analysis is done on old/traditional vision/speech models, and 
their solution relies on offloading some layers to the cloud, which \revi{is counter to} the goal of running inference locally \revi{and efficiently at edge devices}.

\shep{\revi{Concurrent} work on heterogeneous dataflow accelerators (HDA)~\cite{kwon.hpca2021} demonstrates 
the existence of layer-to-layer heterogeneity, and proposes a family of sub-accelerators where each
sub-accelerator has a different dataflow. Our work has two key differences from HDA.
First, while HDA focuses on changing accelerator dataflows, Mensa holistically considers all aspects of accelerator design
(PE array, dataflow, on-chip memory, off-chip memory bandwidth) across a wider variety of state-of-the-art models. 
For example, with HDA, Transducers would still \revi{likely} suffer from high underutilization (\revi{due to} inadequate off-chip bandwidth)
and very low energy efficiency (\revi{due to} high off-chip parameter off-chip traffic and on-chip buffer inefficiency),
while Mensa overcomes these issues.
Second, HDA focuses on optimizing concurrent multi-NN execution, \revi{and does not consider the area overheads and} energy efficiency \revi{goals that need to be met for edge inference (which \titleShort takes into account)}.}


\section{Conclusion}
\label{sec:conclusion}

\sg{We conduct the first \revi{comprehensive workload and} bottleneck analysis of the
Google Edge TPU, a state-of-the-art ML inference accelerator,
\revi{using} 24 state-of-the-art Google edge NN models \revi{including CNNs, LSTMs, Transducers, and RCNNs}. Our analysis reveals that the Edge TPU's monolithic design leads to
significant underutilization and poor energy efficiency for edge NN models, which exhibit significant variation \revi{across and within NN layers}.
We propose \revi{(i)~}a new \revi{HW/SW} framework called \titleShort, \revi{which integrates and manages layer execution across} multiple small heterogeneous accelerators,
each \revi{tailored} to specific layer characteristics; and \revi{(ii)~a runtime \revi{scheduler} for \titleShort to determine which of these
accelerators should execute \revi{which layer}.} Using our \revi{novel observation} that layers \revi{from various NN models} group into
a small number of clusters, we create a \titleShort design
for the Google edge NN models consisting of three
\revi{new} accelerators\revi{, each of which has customized compute, memory, and dataflow characteristics}. \revi{Compared to the Edge TPU, our design improves energy efficiency and throughput by 3.0x and 3.1x, while reducing inference latency by 1.96x, for the 24 state-of-the-art Google Edge NN models. Compared to a state-of-the-art
reconfigurable ML accelerator (Eyeriss v2), our design improves energy efficiency by 2.4x and
throughput by 4.3x for the 24 Google Edge NN models.} \revi{We hope that \titleShort inspires future work on NN acceleration to account for the significant heterogeneity that exists in and across NN models, and to adopt flexible frameworks such as \titleShort to efficiently provide heterogeneous acceleration for both existing and yet-to-be-developed NN model types.}}

\section*{Acknowledgments}
\revi{We thank SAFARI Research Group members for valuable feedback
and the stimulating intellectual environment they provide. We
acknowledge the generous gifts of our industrial partners, especially
Google, Huawei, Intel, Microsoft, and VMware. This research was
partially supported by the Semiconductor Research Corporation.}



{
\bibliographystyle{IEEEtranS}
\bibliography{references}
}


\end{document}